\documentclass{article}

\usepackage{amsmath,amssymb,amsthm,bbm,natbib,color,nccmath}
\usepackage[margin=1in]{geometry}

\usepackage{caption}
\usepackage{graphicx}
\usepackage{subcaption}
\usepackage{afterpage}
\usepackage[hang,flushmargin]{footmisc}
\usepackage{float}
\usepackage{enumitem,kantlipsum}
\usepackage{multirow}
\usepackage[ruled,vlined]{algorithm2e}
\usepackage{tikz}
\usepackage{bm}
\usepackage{changepage}
\usepackage[
  colorlinks,
  citecolor=blue,
  linkcolor=black,
  anchorcolor=red,
  urlcolor=blue
]{hyperref}    

\newtheorem{theorem}{Theorem}
\newtheorem{lemma}{Lemma}
\newtheorem{proposition}{Proposition}

\newtheorem{remark}{Remark}
\newtheorem{definition}{Definition}
\newtheorem{corollary}{Corollary}

\allowdisplaybreaks

\newcommand{\dt}{\mathcal{D}_{\textnormal{train}}}
\newcommand{\dc}{\mathcal{D}_{\textnormal{cal}}}

\newcommand{\R}{\mathbb{R}}

\newcommand{\eps}{\epsilon}
\newcommand{\EE}[1]{\mathbb{E}\left[{#1}\right]}

\newcommand{\Ep}[2]{\mathbb{E}_{{#1}}\left[{#2}\right]}

\newcommand{\PP}[1]{\mathbb{P}\left\{{#1}\right\}}

\newcommand{\One}[1]{{\mathbbm{1}}\left\{{#1}\right\}}

\newcommand{\X}{\mathcal{X}}
\newcommand{\Y}{\mathcal{Y}}
\newcommand{\FWER}{\textnormal{FWER}}
\newcommand{\FDR}{\textnormal{FDR}}
\newcommand{\FDP}{\textnormal{FDP}}
\newcommand{\risk}{\textnormal{Risk}}
\newcommand{\ebh}{\textnormal{e-BH}}

\newcommand{\mh}{\hat{\mu}}

\newcommand{\ch}{\widehat{C}}
\renewcommand{\hat}{\widehat}
\definecolor{ForestGreen}{RGB}{34,139,34}

\usepackage{authblk}
\usetikzlibrary{decorations.pathreplacing}
\allowdisplaybreaks

\title{Conformal ordering: Distribution-free predictive inference for\\ orders and ranks}
\author{Yonghoon Lee}
\author{Zhimei Ren}
\affil{Department of Statistics and Data Science, the Wharton School,\\ University of Pennsylvania}

\date{\today}

\begin{document}

\maketitle

\begin{abstract}
We study distribution-free predictive inference for the ordering and ranking of multiple unobserved outcomes. Although numerous methods, such as conformal prediction, have been developed for prediction and predictive inference of unobserved outcomes, extending them to provide inference for the ordering of multiple outcomes with an interpretable guarantee is not straightforward. To this end, we develop a framework and methodology for uncertainty quantification in predictive ordering.

Our approach begins by reformulating the problem of determining an ordering among multiple objects as a graph inference problem. Our method adopts a multiple-testing-inspired approach and constructs an \emph{uncertainty graph}, where each edge represents a confidently inferred pairwise ordering. We develop a distribution-free procedure for constructing a consistent uncertainty graph while controlling 
either the family-wise error rate or the false discovery rate. Moreover, we show that 
the resulting graphs naturally yield risk-controlling prediction intervals for the ranks of the outcomes. Empirical results demonstrate that the proposed procedure achieves the target guarantees tightly while maintaining reasonable power and prediction-interval widths.
\end{abstract}


\sloppy

\section{Introduction}

Consider a predictive inference problem in which we observe samples $(X_1,Y_1), \ldots, (X_n,Y_n) \in \X \times \R$ and are subsequently given new feature inputs $X_{n+1}, \ldots, X_{n+m}$. The objective is to conduct inference and make decisions regarding the unobserved outcomes $Y_{n+1}, \ldots, Y_{n+m}$. In such settings, the target of inference may vary depending on the application and the goal of the analysis. For example, in clinical trials, one may wish to identify the patients who are most likely to benefit from a treatment, while in hiring or admissions, one may wish to rank applicants according to their predicted future performance and select the most promising candidates. In this work, we study the problem of making distribution-free inference for the ordering of the unobserved outcomes.

Distribution-free predictive inference has mostly been studied in the form of prediction sets, as exemplified by conformal prediction~\citep{vovk2005algorithmic}. Conformal prediction provides prediction sets with valid finite-sample coverage guarantees under the sole assumption of exchangeability of the data points. However, when it comes to downstream tasks such as ordering or ranking multiple unobserved outcomes, it becomes unclear what these guarantees imply about the quality of the resulting decisions. For example, suppose we construct conformal prediction sets for each of the unobserved outcomes $Y_{n+1}, \ldots, Y_{n+m}$ and then apply a decision rule based on these prediction sets to conclude that certain individuals are likely to have the highest outcome values. In such cases, it is not immediately clear how marginal coverage guarantees can be translated into an evaluation of, or uncertainty quantification for, the resulting ordering.

In this work, we develop a procedure that addresses the problem of predictive inference for the ordering and ranks of unobserved outcomes, providing a guarantee that directly quantifies the quality of the ordering. Motivated by {\em Conformal Selection}~\citep{jin2023selection}, we introduce a multiple-testing-type formulation of the predictive ordering problem and a procedure with false discovery rate control, whose interpretation in this context is discussed later. We also show that this procedure leads to risk-controlling simultaneous prediction sets for the ranks.

Regarding the development of an inference procedure that quantifies the uncertainty in the estimated ordering, the main question that needs to be addressed is:
\[\textit{What is a useful form of inference for predictive ordering, and how can we achieve it?}\]
In this work, we introduce an inferential target for predictive ordering that admits a concise interpretation, and propose a procedure that achieves the target guarantee in a distribution-free manner. Throughout the paper, we assume that the data points $(X_1,Y_1), \ldots, (X_n,Y_n), (X_{n+1},Y_{n+1}), \ldots, (X_{n+m},Y_{n+m})$ are exchangeable, but impose no assumptions on their underlying distribution.

\paragraph{Notations.}
We use $\R$ to denote the real space and $\R^+$ to denote the set of nonnegative real numbers. 
For a positive integer $n$, $[n]$ denotes the set $\{1,2,\cdots,n\}$ and $(a_i)_{i \in [n]}$ denotes the vector $(a_1,a_2,\cdots,a_n)$. 
We slightly abuse notation and write $(a_i)_{i \in [n]}$ to also denote the set $\{a_1,a_2,\cdots,a_n\}$ when the meaning is clear. We denote the maximum of the real numbers $a$ and $b$ by $a \vee b$,
and the minimum of the real numbers $a$ and $b$ by $a \wedge b$.
Let $\mathcal{S}_m$ denote the set of all permutations of $[m]$.
\subsection{Main contributions}

We study predictive inference for the ordering of multiple unobserved outcomes. Our contributions are as follows.

\paragraph{A framework for predictive inference for the order.}
We introduce a novel framework for predictive inference on the order of test outcomes. Specifically, we propose an uncertainty quantification framework in the form of a \textit{consistent vertex-ordered uncertainty graph}. We then formulate the problem as a ``multiple testing problem with a random set of random hypotheses"---which introduces additional challenges beyond those in standard multiple testing problems---and develop methods for this inference task.

\paragraph{Uncertainty graph with error control.}
We introduce {\em conformal ordering}, a set of methods for constructing a consistent uncertainty graph with the
family-wise error rate (FWER) or the 
false discovery rate (FDR) control. 
For FWER control, we develop two methods: a Bonferroni-type procedure and a 
method inspired by Tukey's honest significant difference test~\Citep{tukey1949comparing}.
We also provide a complementary hardness result, showing the necessity of having 
sufficiently many calibration samples to achieve non-trivial power.
For FDR control, 
our method applies the e-BH procedure~\citep{wang2022false}, together with the development of an efficient conformal e-value for predictive pairwise comparisons.

\paragraph{Simultaneous prediction intervals for the ranks with risk control.}

We show that a consistent uncertainty graph for ordering with 
FWER or FDR control induces simultaneous prediction intervals 
for the ranks of the outcomes, with control of respective risk measures.

\paragraph{Empirical evaluation.}

We evaluate the performance of the proposed methods on both simulated and real data.\footnote{Code to reproduce the experiments is available at \url{https://github.com/yhoon31/order}.} Empirical results show that the methods achieve their target guarantees---FDR control and risk control---tightly, while maintaining sufficient power and reasonable prediction interval widths.

\subsection{Related work}
Conformal prediction was introduced and studied in the foundational works of~\citet{vovk2005algorithmic} and~\citet{papadopoulos2008inductive}, and has since gained considerable attention in the recent literature on distribution-free inference. For overviews of results in this field, see~\citet{vovk2005algorithmic},~\citet{angelopoulos2021gentle}, and~\citet{angelopoulos2024theoretical}.

Several works have extended the idea of conformal prediction and studied inference methods beyond prediction sets. A representative example closely related to this work is conformal selection, introduced by~\citet{jin2023selection}, in which the authors propose a new framework for predictive inference based on a multiple testing approach. Their work extends ideas from outlier detection~\citet{10.1214/22-AOS2244}, and these ideas have subsequently been extended in~\citet{jin2025model,bashari2024derandomized,marandon2024adaptive,gui2024conformal,liang2024integrative, gui2024conformal}, and~\citet{lee2025selectionhierarchicaldataconformal}.
See also~\citet{lee2024batch} for an alternative selection-type inference.

Similarly to conformal selection, our work is closely related to multiple testing. We introduce a method with false discovery rate control and apply the e-BH procedure~\citep{wang2022false} to achieve the target guarantee. The e-BH procedure is an extension of the Benjamini--Hochberg procedure~\citep{benjamini1995controlling} that operates on e-values instead of p-values. E-values have gained significant interest in the recent literature, and an overview of this area is provided in~\citet{ramdas2025hypothesis}. See, e.g.,~\citet{ignatiadis2024values,fischer2024online,ignatiadis2024compound,ren2024derandomised,lee2024boosting,xu2025bringing,goeman2025partitioning} for recent developments on testing with e-values.

Several works have studied uncertainty quantification methods for classical ranking problems with fixed parameters. \citet{neuhof2024confident} studies uncertainty quantification for the ordering of feature importance, and~\citet{al2022simultaneous} discuss methods for inference on the ranks of means from normal samples. \citet{hung2019rank} developed a valid procedure for rank verification in exponential-family models, and \citet{goldwasser2025gaussian} extended this framework to Gaussian means with unequal variances. \citet{sood2025powerful} further extended rank verification to multivariate Gaussian data with arbitrary covariance structures and to verification of the top-K set.

In the context of predictive inference concerning the ranking of multiple random variables, \citet{fermanian2025transductive} considers the problem of constructing prediction intervals for the rank of a test point among the merged calibration and test data---a problem that differs from that considered in this work---, and proposes methods based on concentration inequalities.
\cite{neuhof2026rank} studies prediction interval for the performance rank 
of models on a new task, assuming access to performance measurements on previous tasks
that are exchangeable with the new task, a setting that differs from ours.

\section{Preliminaries}
Before introducing the proposed methods and their theoretical properties, we provide an overview of the problem formulation and our approaches.

Given training data $\{(X_i,Y_i)\}_{i\in [n]}$ and test features $\{X_{n+j}\}_{j\in [m]}$,
our goal is to construct an order predictor $\hat I \in \mathcal{S}_m$, such that 
\begin{align*}
Y_{n+\hat{I}_1} \leq Y_{n+\hat{I}_2} \leq \cdots \leq Y_{n+\hat{I}_m}
\end{align*}
holds approximately, in a sense to be formalized shortly. In other words, 
$\hat I$ can be viewed as an estimator of the true random {\em ordering}
$I \in \mathcal{S}_m$, defined by $Y_{n+I_1} \leq Y_{n+I_2} \leq \cdots \leq Y_{n+I_m}$. 
When the outcomes are almost surely distinct, 
so that distinct ranks are well-defined almost surely, 
we also consider inference on the {\em ranks} $R_j = \sum_{l=1}^m \One{Y_{n+j} \geq Y_{n+l}}$ for $j \in [m]$.

Since the test outcomes are unobserved, 
a natural approach is to learn a prediction function 
$\hat \mu: \mathcal{X} \mapsto \mathbb{R}$---for example, 
by fitting a regression on a separate training set or a data split---and  
obtain $\hat I$ by ranking the test units according to $\hat \mu (X_{n+j})$.
More specifically, assuming that there are no ties between $\hat{\mu}(X_{n+j})$ values, 
the resulting estimator $\hat{I}$ from this approach can be written as
\begin{equation}\label{eqn:est}
    \hat{I} = (\hat{\sigma}^{-1}(1), \cdots, \hat{\sigma}^{-1}(m)), \text{ where } \hat{\sigma} : [m] \rightarrow [m], ~\hat{\sigma}(j) =  \sum_{k=1}^m \One{\hat{\mu}(X_{n+k}) \leq \hat{\mu}(X_{n+j})} \text{ for } j \in [m].
\end{equation}

For example, if $m=4$ and $\hat{\mu}(X_{n+3}) < \hat{\mu}(X_{n+1}) < \hat{\mu}(X_{n+2}) < \hat{\mu}(X_{n+4})$, then $\hat{I} = (3,1,2,4)$, 
corresponding to the order prediction $Y_{n+3} \leq Y_{n+1} \leq Y_{n+2} \leq Y_{n+4}$.
While $\hat I$ provides a point prediction for true ordering $I$, 
it remains to quantify the uncertainty associated with this prediction. 
In this work, we study how to quantify the uncertainty in the predicted ordering of outcomes.

Throughout the paper, we restrict attention to the setting where 
the outcomes are almost surely distinct, i.e., $\PP{Y_i = Y_j} = 0$,
$\forall i\neq j$.
We also assume that the predicted values $(\hat{\mu}(X_{n+j}))_{j \in [m]}$ are almost surely distinct. This condition can generally be ensured by adding negligible continuous random noise to the predictions from an original predictor $\hat{\mu}_0$, yielding a perturbed predictor $\hat{\mu}$.

\subsection{Graphical representation of the ordering relation}

We begin by formulating the statistical inference problem through a graphical representation 
of the ordering relation.
\begin{proposition}
There is a one-to-one correspondence between the set of all orderings of $m$ elements and the set of all collections of pairwise orderings of these $m$ elements that satisfy transitivity.
\end{proposition}
This result follows from a fundamental result in algorithm theory: any directed acyclic graph admits a topological ordering~\citep{cormen2022introduction}---noting that any set of pairwise orderings with transitivity naturally represents a directed acyclic graph.
For example, in the case of $m=4$, the ordering $(3,1,2,4)$ uniquely determines the set of pairwise orderings $\{(3,1), (3,2), (3,4), (1,2), (1,4), (2,4)\}$, which corresponds to the relations $Y_{n+3} \leq Y_{n+1}$, $Y_{n+3} \leq Y_{n+2}$, $\dots$, $Y_{n+2} \leq Y_{n+4}$. These pairwise orderings satisfy transitivity; for instance, if $Y_{n+3} \leq Y_{n+1}$ and $Y_{n+1} \leq Y_{n+4}$ hold, then $Y_{n+3} \leq Y_{n+4}$ must also hold. Conversely, any set of pairwise orderings of $m$ elements that satisfies transitivity corresponds to a full ordering of $m$ elements. 

Based on this one-to-one correspondence, the task of predicting the total ordering can be reformulated as that of predicting the pairwise orderings. We therefore focus on quantifying the uncertainty in 
these pairwise comparisons.
To be concrete, we aim to construct a procedure that outputs 
\begin{align*}
(\phi_{\hat{I}_{k_1}, \hat{I}_{k_2}})_{1 \leq k_1 < k_2 \leq m}, \quad\text{where each }
\phi_{\hat{I}_{k_1}, \hat{I}_{k_2}} \in \{0,1\}.
\end{align*}
Here, $\phi_{\hat{I}_{k_1}, \hat{I}_{k_2}} = 1$ indicates that 
we certify the ordering $Y_{n+\hat{I}_{k_1}} < Y_{n+\hat{I}_{k_2}}$, 
while $\phi_{\hat{I}_{k_1}, \hat{I}_{k_2}} = 0$ indicates this pairwise comparison remains uncertain,
despite $\hat{\mu}(X_{n+\hat{I}_{k_1}}) < \hat{\mu}(X_{n+\hat{I}_{k_2}})$. 
That is, we identify the pairwise orderings we can certify
after constructing the point estimate of the ordering. 

We note that this is not merely a mathematical formulation. Ordering and ranking 
themselves are essentially about pairwise comparisons---e.g., an element of rank 3 among five objects means that two of the others are larger and the remaining two are smaller. We also note that inference on ranks based on pairwise comparisons has previously appeared in~\citet{neuhof2024confident} in a different context---ordering the means of samples. However, for predictive inference in which the ordering targets are random variables, additional challenges arise, as discussed further in the next section.

Figure~\ref{fig:graph} provides a visual representation of this problem formulation. Our task is to construct an ``uncertainty graph" with ordered vertices, where the solid edges indicate that the corresponding vertices are certified---likely ordered correctly---, 
and the dotted edges imply that we are uncertain about the order of the corresponding outcomes. 
In the next section, we introduce a method for constructing such graphs with a useful inferential guarantee, which we describe in detail.

\begin{figure}[ht]
    \centering
\begin{tikzpicture}

\draw[very thick] (1,0) -- (12,0);

\foreach \x/\label in {2/\text{$\hat{\mu}(X_{n+3})$}, 4/\text{$\hat{\mu}(X_{n+1})$}, 8/\text{$\hat{\mu}(X_{n+2})$}, 11/\text{$\hat{\mu}(X_{n+4})$}} {
    \draw[semithick] (\x,0) -- (\x,-0.3);
    \node at (\x, -0.5) {\label};
}

\draw[dotted, very thick] (2, 0) .. controls (3, 0.3) .. (4, 0);

\draw[very thick] (4, 0) .. controls (6, 0.5) .. (8, 0);

\draw[dotted, very thick] (8, 0) .. controls (9.5, 0.4) .. (11, 0);

\draw[very thick] (2, 0) .. controls (5, 1.4) .. (8, 0);

\draw[very thick] (4, 0) .. controls (7.5, 1.5) .. (11, 0);

\draw[very thick] (2, 0) .. controls (6.5, 3) .. (11, 0);

\node[anchor=west] at (12.5, 1.2) {
    \begin{tabular}{l}
        \textbf{Predicted order:} \\[0.2em]
        $Y_{n+3} \leq Y_{n+1} \leq Y_{n+2} \leq Y_{n+4}$ \\[1.5em]
        \textbf{Pairs for which we have} \\ \textbf{certainty in the order:} \\[0.2em]
        $Y_{n+3} < Y_{n+2}$, $Y_{n+3} < Y_{n+1}$, \\
        $Y_{n+2} < Y_{n+4}$, $Y_{n+1} < Y_{n+4}$
    \end{tabular}
};

\end{tikzpicture}
\caption{A visual representation of the formulation of inference, illustrated with an example where the test size is $m=4$.}
\label{fig:graph}
\end{figure}

We will also see that having an uncertainty graph induces another form of inference, namely, simultaneous prediction intervals for the ranks.

\subsection{Error control on uncertainty graphs}
The graphical representation of the ranking problem gives rise to 
a family of pairwise tests: for any $k_1 < k_2$, we test the hypothesis 
induced by $\hat I_{k_1}$ and $\hat I_{k_2}$:
\begin{align}
\label{eq:null}
H^*_{\hat I_{k_1}, \hat I_{k_2}}: Y_{n+\hat I_{k_1}} \ge Y_{n+ \hat I_{k_2}} .
\end{align}
Therefore, the decision $\phi_{\hat I_{k_1}, \hat I_{k_2}} = 1$ 
corresponds to rejecting $H_{\hat I_{k_1},\hat I_{k_2}}$.
Unlike classical hypotheses about fixed population parameters, 
these hypotheses concern random outcomes, so their truth values
are themselves random.
Related random hypotheses were considered in~\citet{jin2023selection,jin2023model},
who test $Y_{n+j} \le c$ for each $j\in[m]$, where $c\in\mathbb{R}$ is a predifined threshold. 
\citet{lee2025selectionhierarchicaldataconformal} further allow 
the number of the (random) hypotheses to be random.
In our setting, the data-dependent ranking also determines the direction of 
of each pairwise comparison, introducing another source of randomness 
that must be accounted for during inference.

Given the training data $\{(X_i,Y_i)\}_{i\in[n]}$ and the test features $(X_{n+j})_{j\in[m]}$,  a procedure returns uncertainty graph 
represented by
an estimated ordering 
$\hat I \in \mathcal{S}_m$ and 
pairwise certainty indicators: $(\phi_{\hat{I}_{k_1}, \hat{I}_{k_2}})_{1 \leq k_1 < k_2 \leq m} \in \{0,1\}^{m(m-1)/2}$.
We require these indicators to satisfy the following 
{\em consistency} condition with respect to the ordering.

\begin{definition}[Consistency]\label{def:consistency}
Given an estimated ordering $\hat I$, the pairwise indicators
$(\phi_{\hat{I}_{k_1}, \hat{I}_{k_2}})_{1 \leq k_1 < k_2 \leq m}$
are consistent if
$\phi_{\widehat I_{k_1},\widehat I_{k_2}}
\le
\phi_{\widehat I_{k'_1},\widehat I_{k'_2}}$
for all $1\le k'_1\le k_1<k_2\le k'_2\le m$.
\end{definition}
Consistency ensures that the certainty of a comparison is preserved when the comparison is extended outward in the predicted ordering: the first outcome is replaced by one appearing earlier in the ordering, the second by one appearing later, or both. For instance, in the example above with $m=4$ and $\hat{I} = (3,1,2,4)$, 
certifying $Y_{n+1} < Y_{n+2}$ also implies certifying 
$Y_{n+3} < Y_{n+2}$, $Y_{n+1} < Y_{n+4}$, and $Y_{n+3} < Y_{n+4}$.

In addition to consistency, we seek to control false specifications. 
We consider two error metrics:
the false discovery rate (FDR):
\begin{equation}\label{eqn:fdr}
        \FDR = \EE{\frac{\sum_{1 \leq k_1 < k_2 \leq m} \One{\phi_{\hat{I}_{k_1}, \hat{I}_{k_2}} = 1 \text{ and } Y_{n+\hat{I}_{k_1}} \geq Y_{n+\hat{I}_{k_2}}}}{\sum_{1 \leq k_1 < k_2 \leq m} \One{\phi_{\hat{I}_{k_1}, \hat{I}_{k_2}} = 1}\vee 1}},
\end{equation}
and the family-wise error rate (FWER): 
\begin{align}
\FWER = \PP{\exists\, 1\le k_1 <k_2 \le m \text{ such that } \phi_{\hat I_{k_1},\hat I_{k_2}} =1,
Y_{n+\hat I_{k_1}} \ge Y_{n+\hat I_{k_2}}}.
\end{align}

FDR measures the expected proportion of incorrect certifications among all certified pairs, whereas FWER measures the probability of making at least one incorrect certification.
At the same nominal level, FWER control provides a stronger guarantee and therefore generally yields fewer certifications.

Although we adopt standard terminology of multiple testing, we note that our inferential 
targets are pairwise relations among random test objects, 
rather than fixed hypotheses of the data-distribution process. 
Here, a ``false discovery'' corresponds to an incorrect
certification. The expectation in definition~\eqref{eqn:fdr} 
is taken with respect to 
all the training and test data, including the randomness in the $\hat{I}_{j}$'s.

\subsection{Graph-induced prediction intervals for ranks}
Having formulated uncertainty quantification for ordering as a multiple-testing problem and specified the corresponding error metrics, we next discuss how the resulting pairwise certainty indicators can be translated into prediction intervals for the ranks.
Specifically, for the true relative ranks $R_1,\ldots,R_m$ of the test outcomes $Y_{n+1},\ldots,Y_{n+m}$, we construct prediction intervals $\ch_1,\ldots,\ch_m \subseteq [m]$ that aim to cover the true ranks with controlled risk according to the chosen error metric (FWER or FDR).

\paragraph{FWER-induced prediction intervals.}
When the uncertainty graph is constructed to control the FWER at level $\alpha$, 
the target risk is the probability of having at least one prediction interval that fails to cover the true rank, i.e.,
\begin{equation}
    \PP{\exists j \in [m]: R_j \notin \ch_j} \leq \alpha.
\end{equation}

\paragraph{FDR-induced prediction intervals.}
When the uncertainty graph is constructed to control the FDR at level $\alpha$, 
the target risk is defined as a weighted average of the distances between the prediction intervals and the corresponding true ranks. 
\begin{equation}\label{eqn:rank_risk_rough}
    \EE{\sum_{j=1}^m w_j \cdot d(R_j,\ch_j)} \leq \alpha,
\end{equation}
where the weights $w_j$ and the distance measure $d$ are specified in the main results.

The guarantee of the form~\eqref{eqn:rank_risk_rough} is first studied in \citet{benjamini2026uncertainty}
and can be viewed as a generalized version of the standard coverage guarantee, which takes the form of a 0--1 loss. For example, if $d(R_j,\ch_j) = \One{R_j \notin \ch_j}$ and $w_j = 1/m$ for all $j$, then the quantity in~\eqref{eqn:rank_risk_rough} reduces to the average miscoverage rate. While we provide this example for illustration, we also clarify that our procedure establishes the guarantee~\eqref{eqn:rank_risk_rough} for a specific choice of the weights $w_j$ and the distance measure $d$ induced by the uncertainty graph under FDR control, rather than providing a method that treats $w_j$ and $d$ as hyperparameters. As in the construction of the uncertainty graph, the user only needs to choose the level $\alpha$.

\subsection{Interpreting uncertainty graphs through ordered clusters}
We also consider a cluster-based representation as an alternative way to interpret the ordering information provided by the uncertainty graph. A consistent uncertainty graph for ordering naturally induces clusters of adjacent outcomes whose relative ordering is uncertain, suggesting that outcomes within the same cluster tend to have more similar values than those in different clusters. Specifically, the ordered vertices in the uncertainty graph can be grouped into an ordered sequence of clusters, such that the ordering between clusters is inferred to be certain based on the constructed uncertainty graph, as no dotted edge connects vertices from different clusters.

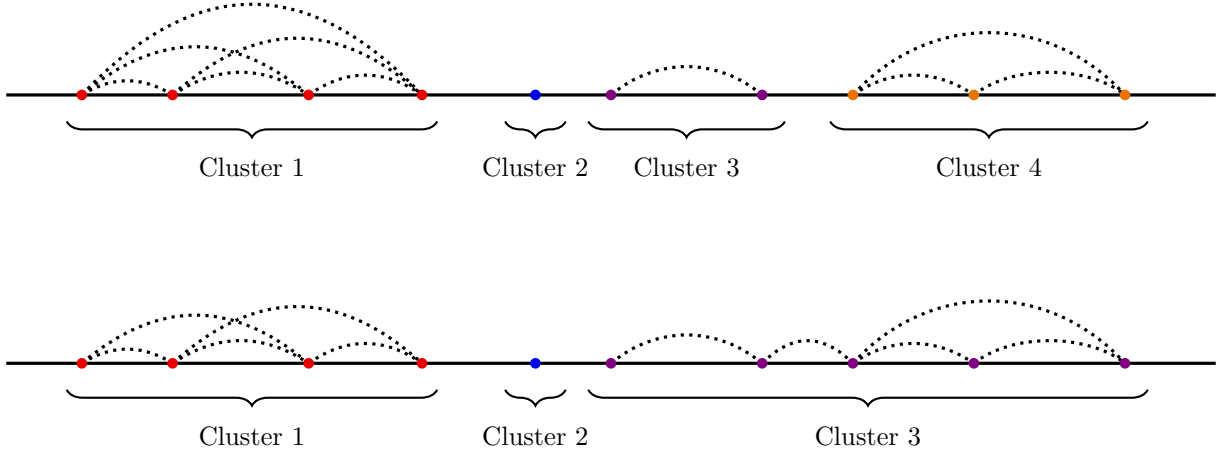
\begin{figure}[ht]
\centering

\begin{tikzpicture}

\draw[very thick] (0,0) -- (16,0);

\draw[dotted, very thick] (1.0,0) .. controls (1.4,0.25) and (1.8,0.25) .. (2.2,0);
\draw[dotted, very thick] (1.0,0) .. controls (2.0,0.85) and (3.0,0.85) .. (4.0,0);
\draw[dotted, very thick] (1.0,0) .. controls (2.5,1.6) and (4.0,1.6) .. (5.5,0);
\draw[dotted, very thick] (2.2,0) .. controls (2.8,0.4) and (3.4,0.4) .. (4.0,0);
\draw[dotted, very thick] (2.2,0) .. controls (3.3,1.0) and (4.4,1.0) .. (5.5,0);
\draw[dotted, very thick] (4.0,0) .. controls (4.5,0.35) and (5.0,0.35) .. (5.5,0);

\draw[dotted, very thick] (8.0,0) .. controls (8.6,0.5) and (9.4,0.5) .. (10.0,0);

\draw[dotted, very thick] (11.2,0) .. controls (11.7,0.35) and (12.3,0.35) .. (12.8,0);
\draw[dotted, very thick] (12.8,0) .. controls (13.3,0.4) and (14.3,0.4) .. (14.8,0);
\draw[dotted, very thick] (11.2,0) .. controls (12.2,1.1) and (13.8,1.1) .. (14.8,0);

\foreach \x in {1.0,2.2,4.0,5.5} {
    \fill[red!90!black] (\x,0) circle (2pt);
}
\fill[blue!90!black] (7.0,0) circle (2pt);
\foreach \x in {8.0,10.0} {
    \fill[violet] (\x,0) circle (2pt);
}
\foreach \x in {11.2,12.8,14.8} {
    \fill[orange!90!black] (\x,0) circle (2pt);
}

\draw[decorate, decoration={brace, mirror, amplitude=6pt}, thick]
    (0.8,-0.35) -- (5.7,-0.35)
    node[midway, below=10pt] {Cluster 1};

\draw[decorate, decoration={brace, mirror, amplitude=6pt}, thick]
    (6.6,-0.35) -- (7.4,-0.35)
    node[midway, below=10pt] {Cluster 2};

\draw[decorate, decoration={brace, mirror, amplitude=6pt}, thick]
    (7.7,-0.35) -- (10.3,-0.35)
    node[midway, below=10pt] {Cluster 3};

\draw[decorate, decoration={brace, mirror, amplitude=6pt}, thick]
    (10.9,-0.35) -- (15.1,-0.35)
    node[midway, below=10pt] {Cluster 4};

\end{tikzpicture}

\vspace{1.2cm}

\begin{tikzpicture}

\draw[very thick] (0,0) -- (16,0);

\draw[dotted, very thick] (1.0,0) .. controls (1.4,0.25) and (1.8,0.25) .. (2.2,0);
\draw[dotted, very thick] (1.0,0) .. controls (2.0,0.85) and (3.0,0.85) .. (4.0,0);
\draw[dotted, very thick] (2.2,0) .. controls (2.8,0.4) and (3.4,0.4) .. (4.0,0);
\draw[dotted, very thick] (2.2,0) .. controls (3.3,1.0) and (4.4,1.0) .. (5.5,0);
\draw[dotted, very thick] (4.0,0) .. controls (4.5,0.35) and (5.0,0.35) .. (5.5,0);

\draw[dotted, very thick] (8.0,0) .. controls (8.6,0.5) and (9.4,0.5) .. (10.0,0);
\draw[dotted, very thick] (10.0,0) .. controls (10.4,0.4) and (10.8,0.4) .. (11.2,0);

\draw[dotted, very thick] (11.2,0) .. controls (11.7,0.35) and (12.3,0.35) .. (12.8,0);
\draw[dotted, very thick] (12.8,0) .. controls (13.3,0.4) and (14.3,0.4) .. (14.8,0);
\draw[dotted, very thick] (11.2,0) .. controls (12.2,1.1) and (13.8,1.1) .. (14.8,0);

\foreach \x in {1.0,2.2,4.0,5.5} {
    \fill[red!90!black] (\x,0) circle (2pt);
}

\fill[blue!90!black] (7.0,0) circle (2pt);

\foreach \x in {8.0,10.0,11.2,12.8,14.8} {
    \fill[violet] (\x,0) circle (2pt);
}

\draw[decorate, decoration={brace, mirror, amplitude=6pt}, thick]
    (0.8,-0.35) -- (5.7,-0.35)
    node[midway, below=10pt] {Cluster 1};

\draw[decorate, decoration={brace, mirror, amplitude=6pt}, thick]
    (6.6,-0.35) -- (7.4,-0.35)
    node[midway, below=10pt] {Cluster 2};

\draw[decorate, decoration={brace, mirror, amplitude=6pt}, thick]
    (7.7,-0.35) -- (15.1,-0.35)
    node[midway, below=10pt] {Cluster 3};

\end{tikzpicture}

\caption{Visual representations of the ordered clusters induced by uncertainty graphs for ordering, with solid edges omitted for simplicity.}
\label{fig:graph_cluster}
\end{figure}

A visual illustration is provided in Figure~\ref{fig:graph_cluster}. The first graph illustrates the case of simple within-cluster structures, where each cluster forms a complete subgraph of dotted edges. The overall ordering structure, together with its associated uncertainty, can then be represented by four ordered clusters $C_1 < C_2 < C_3 < C_4$. In this case, the clusters provide a clear and intuitive grouping of outcomes with similar values.

However, clusters need not correspond to complete subgraphs of dotted edges. The second graph in Figure~\ref{fig:graph_cluster} illustrates such a case: we have three ordered clusters $C_1 < C_2 < C_3$, where $C_1$ and $C_3$ also contain pairs of vertices whose relative ordering are inferred certain. Thus, some certainty information within clusters may be lost when summarizing the uncertainty graph through ordered clusters.

Nevertheless, this cluster-based representation provides an intuitive summary of the overall ordering structure, highlighting the ordering relations between clusters that are inferred to be certain. In practice, it can help identify which groups can be reliably distinguished and where uncertainty in the ordering is concentrated. 

An uncertainty graph with FWER control at level $\alpha$ guarantees that the ordering relations between all clusters are simultaneously correct with probability at least $1-\alpha$, since every pair of vertices from different clusters is connected by a solid edge. In contrast, an uncertainty graph with FDR control does not directly imply a similarly interpretable statistical guarantee for the ordering between clusters---the resulting representation serves more as an intuitive summary of the overall ordering structure.

\subsection{Review of e-values and the e-BH Procedure}
Our strategy for testing the hypotheses in~\eqref{eqn:null_pair_2} is based on e-values~\citep{grunwald2024safe,shafer2021testing,vovk2021values,ramdas2025hypothesis}. E-values are often used as an alternative to p-values in settings where one can benefit from their unique properties, which we discuss below. In standard testing problems with a fixed null hypothesis $H_0$, an \textit{e-variable} is defined as a nonnegative random variable $E$, i.e., $E \geq 0$ a.s., whose expectation under the null is bounded by one: $\Ep{H_0}{E} \leq 1$. A realization $e$ of the e-variable $E$ is called an \textit{e-value}. However, following common convention, we will often also refer to the random variable $E$ itself as an e-value. 

Given an e-variable $E$, a valid test of $H_0$ with type-I error control at level $\alpha$ can be constructed by rejecting whenever $E \geq 1/\alpha$. Now consider the setting of a ``random null event'' $H_0$, which is the setting of interest in this work. In this case, an e-variable is defined as a nonnegative random variable $E$ satisfying 
\begin{equation}\label{eqn:e_variable}
\EE{E \cdot \One{H_0 \text{ is true}}} \leq 1
\end{equation}

In this work, we construct a procedure for testing the multiple null events with false discovery rate control by first designing valid e-values and applying the e-BH procedure~\citep{wang2022false}.

In the standard multiple testing, suppose we have e-values $e_1,e_2,\ldots,e_m$ associated with hypotheses $H_1,H_2,\ldots,H_m$. The e-BH procedure first orders the e-values in descending order, $e_{(1)} \geq \ldots \geq e_{(m)}$, and then rejects the hypotheses corresponding to the largest $k^*$ e-values, where $k^* = \max\{k \in [m] : e_{(k)} \geq m/\alpha k\}$, with $k^* = 0$ if the set is empty. \citet{wang2022false} show that the e-BH procedure controls the FDR at level $\alpha$ as long as the e-values are valid, without requiring any further assumptions on their joint distribution, i.e.,
\[\FDR = \EE{\frac{\sum_{j=1}^m \One{H_j \text{ is true and rejected}}}{\sum_{j=1}^m \One{H_j \text{ is rejected}} \vee 1}} \leq \alpha.\]
In the setting of random $H_1,\ldots,H_m$, the exact same result continues to hold for e-values satisfying~\eqref{eqn:e_variable}, with essentially the same proof. 

\section{Conformal ordering with FWER control}\label{sec:fwer}

In this section, we introduce methods for constructing uncertainty graphs with FWER control, 
as well as the induced prediction intervals for the ranks with simultaneous coverage control.
\subsection{Uncertainty graph with FWER control}

To start, we construct a predictor $\hat{\mu}:\X\to\R$ using a separate training dataset (e.g., constructed via data splitting).
As in~\eqref{eqn:est}, we then obtain a point estimator $\hat{I}$ for the ordering of the unobserved outcomes
by sorting the predicted values $\hat{\mu}(X_{n+j})$ in ascending order.

Next, we observe that the pairwise certification problem in~\eqref{eq:null} 
is equivalent to testing the following $m(m-1)/2$ hypotheses:

\begin{equation}\label{eqn:null_pair_2}
    H_{j_1,j_2} : (Y_{n+j_1}-Y_{n+j_2})\cdot (\hat{\mu}(X_{n+j_1}) - \hat{\mu}(X_{n+j_2})) \le 0, \qquad 1 \leq j_1 < j_2 \leq m.
\end{equation}
In words, rejecting $H_{j_1,j_2}$ certifies that 
$\hat{\mu}$ correctly orders $Y_{n+j_1}$ and $Y_{n+j_2}$. 
We now aim to develop procedures for simultaneously testing this family, whose indices are fixed but 
whose true values depend on the random, unobserved outcomes.

Since the predicted gap $|\hat \mu(X_{n+j_1}) - \hat \mu(X_{n+j_2})|$
assesses the reliability of the predicted ordering, 
it is natural to reject $H_{j_1,j_2}$ when the gap is sufficiently large.
How should we calibrate the rejection threshold? 
We discuss two methods: the Bonferroni procedure based on conformal p-values 
and a partitioning-based method.

\subsubsection{Bonferroni procedure with conformal p-values}

One straightforward approach is to construct a p-value for each test pair 
by comparing its predicted gap with those incorrectly ordered in the calibration set. 
To this end, we may group the calibration observations into $n_1 = \lfloor n/2 \rfloor$
pairs:
\[
(Z_1,Z_2), ~ (Z_3,Z_4), 
~\ldots, ~(Z_{2n_1-1},Z_{2n_1}),
\]
where we write $Z_i = (X_i,Y_i)$ for brevity.
For any $1 \le j_1 < j_2 \le m$, 
the pair $(Z_{n+j_1},Z_{n+j_2})$ is exchangeable with the above calibration pairs. Therefore, following the standard conformal p-value construction, we can define the following valid p-value:
\begin{equation}\label{eqn:p_val}
p_{j_1,j_2} = \frac{\sum_{i=1}^{n_1} \One{|\mh(X_{2i-1}) - \mh(X_{2i})| \ge |\mh(X_{n+j_1}) - \mh(X_{n+j_2})|, (Y_{2i-1} - Y_{2i})(\mh(X_{2i-1}) - \mh(X_{2i})) \le 0} + 1}{n_1 + 1}.
\end{equation}
The following lemma establishes the validity of the p-values 
defined above, whose proof is provided in Appendix~\ref{app:proof-pval}.
\begin{lemma}\label{lem:p_val}
Fix any $1 \le j_1 < j_2 \le m$.
Suppose that $Z_1,\dots,Z_n,Z_{n+j_1}, Z_{n+j_2}$
are jointly exchangeable.
The random variable $p_{j_1,j_2}$ in~\eqref{eqn:p_val} is a valid p-value for $H_{j_1,j_2}$, in the sense that
\[\PP{p_{j_1,j_2} \leq \alpha \textnormal{ and } H_{j_1,j_2} \textnormal{ holds} } \leq \alpha, \qquad \textnormal{for any $\alpha \in (0,1)$}.\]
\end{lemma}
To control the FWER, we apply the Bonferroni procedure to the conformal $p$-values in~\eqref{eqn:p_val}---that is, 
we reject all hypotheses whose corresponding p-values are not greater than $\alpha/\binom{m}{2}$.
The overall procedure is summarized in Algorithm~\ref{alg:fwer} and 
the following theorem shows that this procedure controls the FWER at the desired level and 
is consistent.

\begin{theorem}\label{thm:fwer}
 The ordered uncertainty graph $(\hat{I}, \mathcal{R})$ constructed by Algorithm~\ref{alg:fwer} is consistent and controls the $\FWER$ at level $\alpha$:
    \[\FWER = \PP{\exists\, 1\le k_1 <k_2 \le m \text{ such that } (\hat{I}_{k_1},\hat I_{k_2}) \in \mathcal{R},
Y_{n+\hat I_{k_1}} \ge Y_{n+\hat I_{k_2}}} \leq \alpha.\]
\end{theorem}
The proof is provided in Appendix~\ref{app:proof-fwer}.
We note that, as with the Bonferroni test in standard multiple testing problems, 
this method can lead to conservative inference when the test size is large.
This concern is particularly relevant here since the multiplicity factor $m(m-1)/2$ grows quadratically with the test size: 
the smallest possible p-value is $1/(n_1+1)$, so we need $n_1$ to be at least of the order $m(m-1)/(2\alpha)$ to have a chance of rejecting any hypotheses.
In the simulations, we will see that tight FWER control is attained when the test size is small (relative to the calibration size), while it becomes increasingly conservative as the test size grows.

\begin{remark}
One might consider Holm's procedure as an alternative multiple testing method with $\FWER$ control, but in our setting with random hypotheses, it is not guaranteed to control the $\FWER$ for the notion of $p$-values considered here.
\end{remark}

\begin{algorithm}[ht]
\caption{Conformal ordering with FWER control: Bonferroni's test}
\label{alg:fwer}
\KwIn{Training data $\dt = (X_i',Y_i')_{1 \leq i \leq n_\text{train}}$, Calibration data $\dc = (X_i,Y_i)_{1 \leq i \leq n}$, $s : \X \times \Y \rightarrow \R$, Test inputs $(X_{n+j})_{1 \leq j \leq m}$, Target level $\alpha$.}

\vskip+2ex

{\bf Step 1:} Construct an estimator $\hat{\mu} : \X \rightarrow \R$ using the training data $\dt$.
\vskip 1ex

{\bf Step 2:} Compute $\hat{I}$ according to~\eqref{eqn:est}.

{\bf Step 3:} For each $1 \leq j_1 < j_2 \leq m$, compute the p-value $p_{j_1,j_2}$ accroding to~\eqref{eqn:p_val}.
\vskip 1ex

{\bf Step 4:} (Bonferroni test) Compute the rejection set $\mathcal{R} = \{(j_1,j_2) : p_{j_1,j_2} \leq 2\alpha/m(m-1)\}$

{\bf Output:} Ordered uncertainty graph $(\hat{I},\mathcal{R})$.

\end{algorithm}

\subsubsection{Partitioning-based method}
We now introduce an alternative method, inspired by the approach of~\citet{lee2024batch}. Define the pairwise score $S_{i_1,i_2}$ as
\[S_{i_1,i_2} = |\hat{\mu}(X_{i_1}) - \hat{\mu}(X_{i_2})| \cdot \One{(\hat{\mu}(X_{i_1}) - \hat{\mu}(X_{i_2}))\cdot (Y_{i_1} - Y_{i_2}) \leq 0}, \quad \text{for } 1 \leq i_1 < i_2 \leq n+m.\]
Let $K = \lfloor n/m\rfloor$, and partition the first $Km$ calibration indices into $K$ disjoint sets $G_1,\ldots,G_K$, each of size $m$: $G_k = \{(k-1)m+1,(k-1)m+2,\ldots,km\}$, for $k \in [K]$. We then compute the group-wise scores $S_{G_1}, S_{G_2}, \ldots, S_{G_K}$, defined as the maximum pairwise score within each group:
\[S_{G_k} = \max_{\substack{i_1,i_2\in G_k \\ i_1 < i_2}} S_{i_1,i_2}, \quad k\in[K],\]
and compute the conformal prediction threshold $T$ from these scores, i.e.,
\[T = \text{$\lceil(1-\alpha)(K+1)\rceil$-th smallest value among $S_{G_1},\ldots,S_{G_K}$.}\]
We then construct an uncertainty graph with the edge set $\mathcal{R} = \{(j_1,j_2) : |\hat{\mu}(X_{n+j_1}) - \hat{\mu}(X_{n+j_2})| > T\}$. The overall procedure is summarized in Algorithm~\ref{alg:fwer_par}.
The following theorem shows that the partitioning-based method also controls the $\FWER$ at the desired level
and is consistent.

\begin{theorem}\label{thm:fwer_par}
    The ordered uncertainty graph $(\hat{I}, \mathcal{R})$ constructed by Algorithm~\ref{alg:fwer_par} is consistent and controls the $\FWER$ at level $\alpha$.
\end{theorem}
The proof of Theorem~\ref{thm:fwer_par} is provided in Section~\ref{app:proof-fwer-par}.
The partitioning-based procedure is generally less conservative than the Bonferroni-based procedure,
requiring only  $\left\lfloor \frac{n}{m} \right\rfloor \ge \left\lceil \frac{1}{\alpha} \right\rceil - 1$
for nontrivial inference. The experiments in Section~\ref{sec:fwer_exp} further illustrate this point.

\begin{algorithm}[ht]
\caption{Conformal ordering with FWER control: Partitioning-based method}
\label{alg:fwer_par}
\KwIn{Training data $\dt = (X_i',Y_i')_{1 \leq i \leq n_\text{train}}$, Calibration data $\dc = (X_i,Y_i)_{1 \leq i \leq n}$, Test inputs $(X_{n+j})_{1 \leq j \leq m}$, Target level $\alpha$.}

\vskip+2ex

{\bf Step 1:} Construct an estimator $\hat{\mu} : \X \rightarrow \R$ using the training data $\dt$.
\vskip 1ex

{\bf Step 2:} Compute $\hat{I}$ according to~\eqref{eqn:est}.
\vskip 1ex

{\bf Step 3:} Let $K = \lfloor n/m \rfloor$ and construct $K$ disjoint sets $G_k = \{(k-1)m+1,\ldots,km\}$, for $k \in [K]$.
\vskip 1ex

{\bf Step 4:} For each $k \in [K]$, compute the group-wise score
\[
S_{G_k} = \max_{\substack{i_1,i_2 \in G_k \\ i_1 < i_2}} |\hat{\mu}(X_{i_1})-\hat{\mu}(X_{i_2})| \cdot \One{(\hat{\mu}(X_{i_1})-\hat{\mu}(X_{i_2}))(Y_{i_1}-Y_{i_2}) \leq 0}.
\]
\vskip 1ex

{\bf Step 5:} Compute the threshold $T$ as the $\lceil(1-\alpha)(K+1)\rceil$-th smallest value among $S_{G_1},\ldots,S_{G_K}$, with $T=+\infty$ if $\lceil(1-\alpha)(K+1)\rceil > K$.
\vskip 1ex

{\bf Step 6:} Compute the rejection set
\[
\mathcal{R} = \{(j_1,j_2) : 1 \leq j_1 < j_2 \leq m,\ |\hat{\mu}(X_{n+j_1})-\hat{\mu}(X_{n+j_2})| > T\}.
\]

{\bf Output:} Ordered uncertainty graph $(\hat{I},\mathcal{R})$.

\end{algorithm}

\subsubsection{A hardness result} 
The partitioning-based method reduces the level of multiplicity adjustment compared to the Bonferroni-based method, 
but still requires $n$ to be large relative to $m/\alpha$ for nontrivial inference.
Is this requirement unavoidable? We provide a hardness result below. 

\begin{theorem}\label{thm:hardness}
Fix $n \ge 1$, $m\ge 2$, $\alpha \in (0,1)$, and a prediction function $f: \mathcal{X} \to \mathbb{R}$.
Consider any procedure that outputs a rejection set $\mathcal{R}$ based on 
$n$ training samples $\{(X_i,Y_i)\}_{i=1}^n$ and $m$ test samples $\{X_{n+j}\}_{j=1}^m$, 
possibly depending on external randomization encoded by $U \sim \textnormal{Unif }[0,1]$.
If $\mathcal{R}$ is consistent with respect to $f$ 
and controls the $\FWER$ at level $\alpha$ for 
any exchangeable distribution of $\{(X_i,Y_i)\}_{i=1}^{n+m}$, 
then 
\begin{align*}
\PP{\mathcal{R} = \varnothing} \ge \sum_{r=1}^{n+1} (p_r - \alpha)_+,
\end{align*}
where $p_r = \frac{\binom{n+m-r}{m-1}}{\binom{n+m}{m}}$.
\end{theorem}
For the right-hand side of the inequality, 
note that $p_1 = \binom{n+m-1}{m-1}/\binom{n+m}{m} = \frac{m}{n+m}$, 
and a simplified bound is therefore $\PP{\mathcal{R} = \varnothing} \ge (\frac{m}{n+m} - \alpha)_+$.
In order to obtain nontrivial inference with probability at least $1-\beta$, 
it is necessary that $n \ge \frac{1-\alpha -\beta}{ \alpha+\beta} m$.

\subsection{Simultaneous prediction intervals for the ranks with coverage control}

We now show that constructing an uncertainty graph with FWER control 
induces prediction sets for the ranks with simultaneous coverage control. 
Specifically, we aim to construct prediction sets $\ch_1, \ldots, \ch_m \subseteq [m]$
for the ranks $R_1, \ldots, R_m$ of the test points.
Let $(\hat{I},\mathcal{R})$ be an ordered uncertainty graph with FWER control at level $\alpha$,
obtained---e.g., from Algorithm~\ref{alg:fwer_par}---, and let $\mathcal{R}_{\rm ord}$ denote the set of certified edges expressed in 
the predicted orders:
\begin{align}\label{eq:rord}
\mathcal{R}_{\rm ord} = \big\{(k_1,k_2): 1\le k_1 < k_2 \le m,  
~(\hat I_{k_1}\wedge \hat I_{k_2}, \hat I_{k_1}\vee \hat I_{k_2}) \in \mathcal{R}\big\}. 
\end{align}
We begin by constructing one-sided prediction sets for the ranks. For $k=1,2,\ldots,m$, let
\begin{equation}\label{eqn:rank_interval}
    \tilde{C}_k = \{1,2,\ldots,\hat{u}_k\}, \qquad \text{where } \hat{u}_k = m - |\{l>k : (k,l)\in\mathcal{R}_{\rm ord}\}|.
\end{equation}
Here, $m-\hat u_k$ counts the number of ranks that are certified to be larger than
$Y_{n+\hat I_k}$, and thus 
$\tilde{C}_k$ is the one-sided prediction set for the rank of $Y_{n+\hat I_k}$.
Returning to the original indices, 
we define the prediction set as $\hat C_{\hat I_k} = \tilde{C}_{k}$, $\forall k \in [m]$.
The following theorem establishes the simultaneous coverage guarantee of these prediction sets.

\begin{theorem}\label{thm:rank_interval}
    Suppose that $(\hat{I},\mathcal{R})$ is an ordered uncertainty graph with $\FWER$ control at level $\alpha$, i.e., satisfies
    \[\PP{\sum_{1 \leq j_1 < j_2 \leq m} \One{H_{j_1,j_2} \textnormal{ holds but } (j_1,j_2) \in \mathcal{R}} > 0} \leq \alpha.\]
    Then the prediction sets $(\ch_j)_{j \in [m]}$ from~\eqref{eqn:rank_interval} satisfy the following coverage guarantee.
    \begin{equation*}\label{eqn:rank_coverage}
    \PP{R_j \in \ch_j \textnormal{ for all } j \in [m]} \geq 1-\alpha.
    \end{equation*}
\end{theorem}

The proof of Theorem~\ref{thm:rank_interval} is straightforward and can be 
found in Appendix~\ref{app:proof-rank-interval};
similar arguments appear in~\citet{neuhof2024confident} and~\citet{al2022simultaneous}.

The two-sided prediction intervals can be derived analogously. 
With $\mathcal{R}$ controlling the FWER at level $\alpha/2$, 
for $k=1,2,\ldots,m$, we let
\begin{equation}\label{eqn:rank_interval_two_sided}
    \tilde{C}_k = \{\hat{l}_k, \hat{l}_k+1,\ldots,\hat{u}_k\}, \; \text{where } \hat{l}_k = 1 + |\{l<k : (l,k)\in\mathcal{R}_{\rm ord}\}| \; \text{and} \; \hat{u}_k = m - |\{l>k : (k,l)\in\mathcal{R}_{\rm ord}\}|.
\end{equation}
We then define the prediction set $\hat C_{\hat I_k} = \tilde{C}_{k}$, $\forall k \in [m]$.

\begin{corollary}\label{cor:rank_fwer_two_sided}
    Suppose that $(\hat{I},\mathcal{R})$ is an ordered uncertainty graph with $\FWER$ control at level $\alpha/2$. Then the prediction sets $(\ch_j)_{j \in [m]}$ from~\eqref{eqn:rank_interval_two_sided} satisfy 
    \begin{equation*}
    \PP{R_j \in \ch_j \textnormal{ for all } j \in [m]} \geq 1-\alpha.
    \end{equation*}
\end{corollary}

We note that, as this method is based on an uncertainty graph with FWER control, which is typically conservative, the resulting prediction intervals also tend to be conservative, i.e., have large widths---especially when the test size is large.

\section{Conformal ordering with FDR control}\label{sec:main_results}
We now construct an uncertainty graph under FDR control, 
as a more liberal alternative to FWER, and derive the corresponding rank prediction intervals.

\subsection{Uncertainty graph construction}

Recall that the conformal p-values~\eqref{eqn:p_val} are individually valid, but they are dependent 
since different p-values share test units and reuse the same 
calibration set. As a result, applying the BH procedure~\citep{benjamini1995controlling} to the p-values do not 
necessarily yield FDR control.

To address this challenge, we design a procedure based on e-values in parallel 
to~\eqref{eqn:p_val}. In particular, 
adapting ideas from~\citet{bashari2023derandomized} and~\citet{lee2025selectionhierarchicaldataconformal}, 
for each $1 \leq j_1 < j_2 \leq m$, we construct e-values for testing $H_{j_1,j_2}$ as 
\begin{equation}\label{eqn:e_val}
    e_{j_1,j_2} = \frac{(n_1+1)\cdot\One{|\hat{\mu}(X_{n+j_1}) - \hat{\mu}(X_{n+j_2})| > T_{j_1,j_2}}}{\sum_{i=1}^{n_1} \One{|\hat{\mu}(X_{2i-1}) - \hat{\mu}(X_{2i})| > T_{j_1,j_2}}\cdot \One{(Y_{2i-1}-Y_{2i})(\hat{\mu}(X_{2i-1}) - \hat{\mu}(X_{2i})) \le 0} + 1} ,
\end{equation}
where
\begin{equation}\label{eqn:fdr_thres}
\begin{split}
    &T_{j_1,j_2} = \inf\Bigg\{t>0 : \widehat{\FDP}_{j_1,j_2}(t) \leq \alpha\Bigg\}, \text{ where }\\
    &\widehat{\FDP}_{j_1,j_2}(t) = \frac{\sum_{i=1}^{n_1} \One{|\hat{\mu}(X_{2i-1}) - \hat{\mu}(X_{2i})| > t}\cdot \One{(Y_{2i-1}-Y_{2i})(\hat{\mu}(X_{2i-1}) - \hat{\mu}(X_{2i})) \le 0} + 1}{\max\bigg\{\sum_{\substack{1\le j_1'<j_2'\le m\\
j_1',j_2'\notin\{j_1,j_2\}}} \One{|\hat{\mu}(X_{n+j_1'}) - \hat{\mu}(X_{n+j_2'})| > t}, 1\bigg\}}\\
&\hspace{130mm}\cdot \frac{(m-2)(m-3)/2}{n_1+1}.
\end{split}
\end{equation}
We let $T_{j_1,j_2} = +\infty$ if there is no $t$ such that $\hat{\FDP}_{j_1,j_2}(t) \leq \alpha$. 
We prove in the following lemma that the e-values are valid, 
with its proof deferred to Appendix~\ref{app:proof-eval}

\begin{lemma}\label{lem:fdr}
    Fix any $1 \le j_1 < j_2 \le m$. Suppose that $Z_1,\dots,Z_{n},Z_{n+j_1},Z_{n+j_2}$ 
    are jointly exchangeable conditional on $\{Z_{n+\ell}\}_{\ell \notin \{j_1,j_2\}}$.
    The random variable $e_{j_1,j_2}$ defined in~\eqref{eqn:e_val} is a valid e-value for $H_{j_1,j_2}$ in~\eqref{eqn:null_pair_2} in the sense that it satisfies $\EE{e_{j_1,j_2} \cdot \One{H_{j_1,j_2} \text{ holds}}} \leq 1$.
\end{lemma}

Since the e-values are valid, applying the e-BH procedure to $\{e_{j_1,j_2}\}_{1\le j_1<j_2\le m}$ yields a rejection set $\mathcal{R}$ that controls the FDR at level $\alpha$.
However, the e-values need not increase with the predicted gaps
so the e-BH rejection set may fail to satisfy the consistency condition in Definition~\ref{def:consistency}. 
To address this issue, we use a more general sufficient condition for FDR control:
a rejection set $\mathcal{R}$ controls the FDR at level $\alpha$ provided 
it is {\em self-consistent}~\citep{wang2022false}, that is, if it satisfies 
\begin{align}\label{eq:self-consistency}
e_{j_i,j_2} \ge \frac{m(m-1)/2}{\alpha |\mathcal{R}|} \quad \text{for all } (j_1,j_2) \in \mathcal{R}.
\end{align}
We then seek a maximum-cardinality rejection set satisfying both the e-value
self-consistency condition in~\eqref{eq:self-consistency} and the consistency condition in Definition~\ref{def:consistency}.
Since all the constraints are algorithmic, 
this task can be formulated as an integer program. 
For any $1\le j_1 < j_2 \le m$, let $\delta_{j_1,j_2} \in \{0,1\}$ denote whether $(j_1,j_2)$ 
is included in the rejection set $\mathcal{R}$. We also write $B_{j_1,j_2} = [\hat \mu_{j_1} \wedge \hat \mu_{j_2}, \hat \mu_{j_1} \vee \hat \mu_{j_2}]$, 
where $\hat \mu_j = \hat{\mu}(X_{n+j})$ for $j \in [m]$. 
The integer program can be formulated as follows:
\begin{equation}
\begin{aligned}\label{eq:lp}
& \max_{\delta_{j_1,j_2} \in \{0,1\}} \sum_{1 \leq j_1 < j_2 \leq m} \delta_{j_1,j_2} \\ 
\text{s.t. } & \alpha \cdot e_{j_1,j_2} \sum_{1 \leq a < b \leq m} \delta_{a,b} \geq 
\delta_{j_1,j_2} \cdot m(m-1)/2, \quad \forall 1 \leq j_1 < j_2 \leq m;  \\
& \delta_{j_1,j_2} \le \delta_{a,b}, \quad \text{ if } 
B_{j_1,j_2} \subseteq B_{a,b},~\forall 1 \leq j_1 < j_2 \leq m, 1\le a < b \leq m.
\end{aligned}  
\end{equation}
Letting $\{\hat \delta_{j_1,j_2}\}_{1 \leq j_1 < j_2 \leq m}$ denote the solution to the integer program in~\eqref{eq:lp}, we define the corresponding rejection set as 
$\mathcal{R} = \{(i,j) : \hat \delta_{j_1,j_2} = 1\}.$
The FDR control guarantee follows directly from the validity of the e-values and the self-consistency of the rejection set $\mathcal{R}$, as formalized in the following theorem.
Its proof is provided in Appendix~\ref{app:proof-fdr}.

\begin{theorem}\label{thm:fdr}
    Given e-values $(e_{j_1,j_2})_{1 \leq j_1 < j_2 \leq m}$ from~\eqref{eqn:e_val}, any rejection set $\mathcal{R}$ that satisfies the 
    self-consistency condition in~\eqref{eq:self-consistency} at level $\alpha$ guarantees $\FDR \leq \alpha$, i.e,
    \[\FDR = \EE{\frac{\sum_{1 \leq j_1 < j_2 \leq m}\One{H_{j_1,j_2} \textnormal{ is true and rejected}}}{\sum_{1 \leq j_1 < j_2 \leq m}\One{H_{j_1,j_2} \textnormal{ is rejected }}\vee 1}} \leq \alpha.\]
\end{theorem}

\paragraph{Computational shortcut.}
The integer program in~\eqref{eq:lp} can be efficiently solved using an 
equivalent e-BH representation. Specifically, for any pair $(j_1,j_2)$ with $1 \leq j_1 < j_2 \leq m$, we define 
\begin{align}\label{eq:shortcut_eval}
\tilde e_{j_1,j_2} = \min_{1 \le a < b \le m, B_{j_1,j_2} \subseteq B_{a,b}  } e_{a,b},
\end{align}
and apply the e-BH procedure to the values $\tilde e_{j_1,j_2}$ at level $\alpha$ to obtain the rejection set $\mathcal{R}_{\text{e-BH}}(\{\tilde{e}_{j_1,j_2}\}_{1 \leq j_1 < j_2 \leq m})$. 
Note that the random variable $\tilde{e}_{j_1,j_2}$ above satisfies $\tilde{e}_{j_1,j_2} \leq e_{j_1,j_2}$ deterministically, and thus it is also a valid e-variable by Lemma~\ref{lem:fdr}.
The proposition below formalizes this equivalence between the e-BH procedure applied to the modified e-values 
and the rejection set obtained from the integer program in~\eqref{eq:lp}.
\begin{proposition}\label{prop:equiv}
The rejection set formed by the solution $\{\hat \delta_{j_1,j_2}\}_{1 \leq j_1 < j_2 \leq m}$ to 
the integer program in~\eqref{eq:lp} is equivalent to the e-BH procedure applied to the modified e-values 
$\tilde e_{j_1,j_2}$, i.e.,
\[
\{(j_1,j_2): \hat{\delta}_{j_1,j_2} = 1\} = \mathcal{R}_{\ebh}(\{\tilde{e}_{j_1,j_2}\}_{1 \leq j_1 < j_2 \leq m})
\]
\end{proposition}
The proof is in Appendix~\ref{app:proof-equiv}, and we summarize the overall procedure in Algorithm~\ref{alg:fdr}. We note that the thresholds $T_{j_1,j_2}$ in~\eqref{eqn:fdr_thres} often have similar values---as they are based on similar FDP estimates $\widehat{\FDP}_{j_1,j_2}(t)$---, making the e-values $e_{j_1,j_2}$ in~\eqref{eqn:e_val} nearly monotone in the predicted gaps. Consequently, applying the e-BH procedure directly to the original e-values is likely to already yield a consistent graph. The adjustment in~\eqref{eq:shortcut_eval} typically changes the e-values only slightly, while ensuring consistency rigorously.

We close this section with the following remark.

\begin{remark}
To improve power or reduce the potential conservativeness of $\FDR$ control, one might ask whether it is possible to effectively reduce the number of hypotheses being tested through a sequential approach, e.g., testing for the adjacent points first in Figure~\ref{fig:graph} and proceeding to the next nearest pairs only for the adjacent points that were not rejected. This idea is based on the intuition that, for ``easy pairs,'' such as $(Y_{n+3},Y_{n+4})$ in Figure~\ref{fig:graph}, we may be able to make decisions more easily and exclude them from the multiplicity adjustment.

However, we note that this is not a viable approach in our setting, which concerns predictive inference rather than classical hypothesis testing. Detecting ``adjacent'' or ``easy'' pairs requires looking at the test points, and thus, once a decision is made based on such information, it breaks the exchangeability of the data points, which is required to derive a procedure with theoretical guarantees.
\end{remark}

\begin{algorithm}[ht]
\caption{Conformal ordering with FDR control}
\label{alg:fdr}
\KwIn{Training data $\dt = (X_i',Y_i')_{1 \leq i \leq n_\text{train}}$, Calibration data $\dc = (X_i,Y_i)_{1 \leq i \leq n}$, $s : \X \times \Y \rightarrow \R$, Test inputs $(X_{n+j})_{1 \leq j \leq m}$, Target FDR level $\alpha$.}

\vskip 1ex

{\bf Step 1:} Construct an estimator $\hat{\mu} : \X \rightarrow \R$ using the training data $\dt$.
\vskip 1ex

{\bf Step 2:} Compute $\hat{I}$ according to~\eqref{eqn:est}.
\vskip 1ex

{\bf Step 3:} For each $1 \leq j_1 < j_2 \leq m$, construct the e-value $e_{j_1,j_2}$ according to~\eqref{eqn:e_val}, using the\\
\hspace{13mm} calibration data $\dc$.
\vskip 1ex

{\bf Step 4:} For each $1 \leq j_1 < j_2 \leq m$, compute the adjusted e-values $\tilde{e}_{j_1,j_2}$ according to~\eqref{eq:shortcut_eval}.
\vskip 1ex

{\bf Step 5:} Sort the e-values $\tilde{e}_{(1)} \geq \tilde{e}_{(2)} \geq \ldots \geq \tilde{e}_{(m(m-1)/2)}$.
\vskip 1ex

{\bf Step 6:} Compute $k^* = \max\Big\{k \in [m(m-1)/2]: \frac{k \tilde{e}_{(k)}}{m(m-1)/2} \geq \frac{1}{\alpha}\Big\}$, 
with $\max \varnothing = 0$.
 
\vskip 1ex

{\bf Step 7:} Compute the rejection set $\mathcal{R} = \{(j_1,j_2) : \tilde{e}_{j_1,j_2} \geq \tilde{e}_{(k^*)}\}$ if 
$k^* > 0$, and $\mathcal{R} = \varnothing$ otherwise.

{\bf Output:} Ordered uncertainty graph $(\hat{I},\mathcal{R})$.

\end{algorithm}

\subsubsection{Improving power with U-eBH procedure}\label{sec:uebh}

In standard multiple testing with e-values, several methods have been developed to improve the power of the e-BH procedure. A representative example is the U-eBH procedure~\citep{xu2023more}, which applies the e-BH procedure after dividing the e-values by an independent uniform random variable $U \sim \mathrm{Unif}([0,1])$. Below, we show that the U-eBH procedure also applies to our setting with random hypotheses.

\begin{corollary}\label{cor:uebh}
    The U-eBH procedure applied to $(\tilde{e}_{j_1,j_2})_{1 \leq j_1 < j_2 \leq m}$ from~\eqref{eq:shortcut_eval} at level $\alpha$ guarantees $\FDR \leq \alpha$.
\end{corollary}

Note that Corollary~\ref{cor:uebh} does not follow directly from the proof of the standard U-eBH procedure, as our problem is 
not a standard multiple testing problem. The validity of the U-eBH procedure in the predictive selection task has been established in~\citet{lee2025selectionhierarchicaldataconformal}, and Corollary~\ref{cor:uebh} follows from similar arguments.

It is worth noting that, in our method, the U-eBH procedure has a limited effect on the composition of the rejection set due to the particular structure of our e-values in~\eqref{eqn:e_val}, which have an indicator in the numerator. Specifically, whether an e-value is positive or zero is determined by whether the corresponding test statistic exceeds its threshold, and dividing the e-values by a uniform random variable $U$ does not change which e-values are positive. Instead, it scales all e-values upward by a common random factor, thereby reducing the likelihood that the e-BH procedure makes no rejections. In Section~\ref{sec:exp_uebh}, we empirically verify that the output of the procedure is indeed stable, i.e., robust to the particular value of $U$ drawn.

As an additional remark, in standard testing problems with fixed hypotheses, there are other methods for improving the power, such as those based on the closure principle~\citep{xu2025bringing,goeman2025partitioning}. However, these approaches do not directly apply to our predictive inference problem with random hypotheses.

\subsection{Simultaneous prediction intervals for ranks with risk control}\label{sec:rank}

We next use the uncertainty graph with FDR control to
construct prediction intervals for the ranks $R_1,\ldots,R_m$.

For one-sided intervals, the overall method given the uncertainty graph $\mathcal{R}$ is identical to that in~\eqref{eqn:rank_interval}. 
However, if $\mathcal{R}$ controls $\FDR$ instead, we can attain the following 
risk control guarantee.

\begin{theorem}\label{thm:rank_fdr}
    Suppose that $(\hat{I},\mathcal{R})$ is an ordered uncertainty graph with $\FDR$ control at level $\alpha$, i.e., satisfies
    \begin{equation}\label{eqn:graph_fdr_guarantee}
        \EE{\sum_{1 \leq j_1 < j_2 \leq m} \One{H_{j_1,j_2} \textnormal{ holds but } (j_1,j_2) \in \mathcal{R}} / \max\{|\mathcal{R}|,1\}} \leq \alpha.
    \end{equation}
    Then the prediction sets $(\ch_j)_{j \in [m]}$ from~\eqref{eqn:rank_interval} satisfy the following guarantee.
    \begin{equation}\label{eqn:rank_target}
    \EE{\frac{\sum_{j=1}^m d_+(R_j,\ch_j)}{\max\{\sum_{k=1}^m (m - \hat{u}_k),1\}}} \leq \alpha,
    \end{equation}
    where $d_+(r,C) = \min_{t \in C} (r-t)_+$ denotes the right-sided distance between $r \in [m]$ and $C \subseteq [m]$.
\end{theorem}

\paragraph{Interpretation of the guarantee}
The term in~\eqref{eqn:rank_target} that is controlled by the procedure measures the extent to which the actual ranks $R_1,\ldots,R_m$ deviate from the prediction intervals $\ch_1,\ldots,\ch_m$. To see this, first observe that $\sum_{k=1}^m (m - \hat{u}_k) = \sum_{j=1}^m (m - \hat{u}_{\hat{I}_j})$, 
and note that the term $m - \hat{u}_{j}$ represents the largest possible value of $d_+(R_{\hat I_j},\ch_{\hat I_j})$, that is, the largest possible distance between $R_{\hat I_j}$ and $\ch_{\hat I_j}$. Thus, the guarantee~\eqref{eqn:rank_target} is equivalent to
\[\EE{\frac{\sum_{j=1}^m d_+(R_j,\ch_j)}{\sum_{j=1}^m \max_{r \in [m]} d_+(r,\ch_j)}} = \EE{\sum_{j=1}^m w_j \cdot \bar{d}_+(r,\ch_j)}\leq \alpha, \]
where
\[w_j = \frac{\max_{r \in [m]} d_+(r,\ch_j)}{\sum_{j=1}^m \max_{r \in [m]} d_+(r,\ch_j)} \quad \text{ and } \quad \bar{d}_+(r,\ch_j) = \frac{d_+(R_j,\ch_j)}{\max_{r \in [m]} d_+(r,\ch_j)},\]
with the convention that $0/0 = 0$.
Therefore, the guarantee~\eqref{eqn:rank_target} can be viewed as controlling a weighted mean of the standardized distance $\bar{d}_+(r,\ch_j)$ between the ranks and the prediction intervals, where the weights reflect the ``maximum possible risk" of failing to cover each individual rank.

\begin{remark}
The idea of constructing prediction sets with a risk control guarantee, rather than standard coverage guarantees, has appeared in~\citet{angelopoulos2022conformal}---where the authors consider the standard predictive inference problem of constructing a prediction set $\ch(X_{n+1})$ for $Y_{n+1}$ and introduce a method with a guarantee of the form
\[\EE{\ell(\ch(X_{n+1}), Y_{n+1})} \leq \alpha,\]
for a predefined loss function $\ell$.
\end{remark}

For two-sided intervals, one can apply the formula in~\eqref{eqn:rank_interval_two_sided}, with $\mathcal{R}$ controlling the $\FDR$ at level $\alpha$.
Similarly to Theorem~\ref{thm:rank_fdr}, we obtain the following guarantee for the two-sided prediction intervals,
whose proof is omitted as it is analogous to that of Theorem~\ref{thm:rank_fdr}.

\begin{corollary}\label{cor:rank_fdr_two_sided}
    Suppose that $(\hat{I},\mathcal{R})$ is an ordered uncertainty graph with $\FDR$ control at level $\alpha$ in the sense of~\eqref{eqn:graph_fdr_guarantee}.
    Then the prediction sets $(\ch_j)_{j \in [m]}$ from~\eqref{eqn:rank_interval_two_sided} satisfy $\max\{\risk+, \risk-\} \leq \alpha$, where
    \begin{equation}\label{eqn:rank_risk}
    \risk+ = \EE{\frac{\sum_{j=1}^m d_+(R_j,\ch_j)}{\sum_{k=1}^m (m - \hat{u}_k)}},\qquad \text{and} \qquad \risk- = \EE{\frac{\sum_{j=1}^m d_-(R_j,\ch_j)}{\sum_{k=1}^m  (\hat{l}_k-1)}}.
    \end{equation}
    Here, $d_+(r,C) = \min_{t \in C} (r-t)_+$ and $d_-(r,C) = \min_{t \in C} (t-r)_+$ denotes the right-sided and left-sided distance between $r \in [m]$ and $C \subseteq [m]$, respectively.
\end{corollary}
The overall procedure is summarized in Algorithm~\ref{alg:rank_fdr}.

\begin{algorithm}[ht]
\caption{Uncertainty graph-induced prediction intervals for the ranks}
\label{alg:rank_fdr}
\KwIn{Training data $\dt = (X_i',Y_i')_{1 \leq i \leq n_\text{train}}$, Calibration data $\dc = (X_i,Y_i)_{1 \leq i \leq n}$, 
Test inputs $(X_{n+j})_{1 \leq j \leq m}$, Target level $\alpha$.}

\vskip 1ex

{\bf Step 1:} Apply Algorithm~\ref{alg:fdr} to obtain the ordered uncertainty graph $(\hat{I}, \mathcal{R})$,
and compute $\mathcal{R}_{\rm ord}$ according to~\eqref{eq:rord}.
\vskip 1ex

{\bf Step 2:} For each $k \in [m]$, compute
$\hat{l}_k = 1 + |\{l<k : (l,k)\in\mathcal{R}_{\rm ord}\}|$ and
$\hat{u}_k = m - |\{l>k : (k,l)\in\mathcal{R}_{\rm ord}\}|$.
\vskip 1ex

{\bf Step 3:} For each $k \in [m]$, construct the prediction interval
$\ch_{\hat I_k} = \{\hat{l}_{k},\hat{l}_{k}+1,\ldots,\hat{u}_{k}\}$.

{\bf Output:} Prediction intervals $\ch_1,\ch_2,\ldots,\ch_m$.

\end{algorithm}

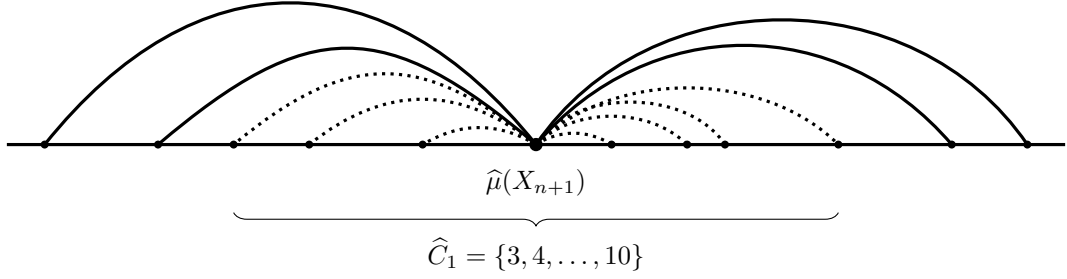
\begin{figure}[ht]
\centering
\begin{tikzpicture}

\draw[very thick] (0,0) -- (14,0);

\foreach \x in {0.5,2,3,4,5.5,7,8,9,9.5,11,12.5,13.5} {
    \fill (\x,0) circle (1.5pt);
}

\fill (7,0) circle (2.5pt);

\coordinate (C) at (7,0);

\draw[very thick, dotted] (C) .. controls (6.5,0.3) and (6,0.3) .. (5.5,0);
\draw[very thick, dotted] (C) .. controls (6,0.8) and (5,0.8) .. (4,0);
\draw[very thick, dotted] (C) .. controls (5.5,1.25) and (4.5,1.25) .. (3,0);
\draw[very thick] (C) .. controls (5,1.7) and (4,1.7) .. (2,0);
\draw[very thick] (C) .. controls (5,2.5) and (2.5,2.5) .. (0.5,0);

\draw[very thick, dotted] (C) .. controls (7.3,0.2) and (7.7,0.2) .. (8,0);
\draw[very thick, dotted] (C) .. controls (7.5,0.5) and (8.5,0.5) .. (9,0);
\draw[very thick, dotted] (C) .. controls (7.5,0.75) and (9,0.75) .. (9.5,0);
\draw[very thick, dotted] (C) .. controls (8,1) and (10,1) .. (11,0);
\draw[very thick] (C) .. controls (8.5,1.75) and (11,1.75) .. (12.5,0);
\draw[very thick] (C) .. controls (8.5,2.2) and (12,2.2) .. (13.5,0);

\node at (7, -0.5) {$\hat{\mu}(X_{n+1})$};

\draw[decorate,decoration={brace,mirror,amplitude=5pt}]
(3,-0.9) -- (11,-0.9)
node[midway, below=6pt] {$\ch_1 = \{3,4,\ldots,10\}$};
\end{tikzpicture}
\caption{Illustration of Procedure~\ref{alg:rank_fdr} with test size $m = 12$. The dots represent the estimated outcome values $\hat{\mu}(X_{n+j})$. The solid lines represent the edges rejected by Procedure~\ref{alg:fdr} associated with $Y_{n+1}$. In this example, the prediction interval for the rank of $Y_{n+1}$ is $\{3,4,\ldots,10\}$.}
\label{fig:rank_interval}
\end{figure}

An illustration of this procedure is provided in Figure~\ref{fig:rank_interval}. For each test point, we consider the point estimate $\hat{\mu}(X_{n+j})$ and the edges in the uncertainty graph that have this point as one of their vertices. As in the representation in Figure~\ref{fig:graph}, we represent the rejected hypotheses as solid lines and the remaining ones as dotted lines. For consistent graphs, the dotted lines are located nearby, while solid lines connect to more distant vertices. The prediction interval is then determined by the number of solid edges connecting each point to vertices on its left and right, which coincides with the interval determined by the farthest dotted connections when the graph is consistent.

\section{Experimental results}

\subsection{Simulations}\label{sec:sim}

We provide simulation results to illustrate the performance of the proposed procedures. We present the results for FDR-controlling procedures here and defer those for $\FWER$-controlling procedures to Appendix~\ref{sec:fwer_exp}.

We consider two settings for data generation and the estimation method.
\begin{align*}
&\textnormal{Setting 1: } X \sim \mathcal{N}_p(\mu, \sigma^2 I_p), \;
Y \mid X \sim \mathcal{N}\bigl(\beta_1^\top X, 1\bigr),\, &&\textnormal{$\hat{\mu}$: from linear regression},\\
&\textnormal{Setting 2: } X \sim \mathcal{N}_p(\mu, \sigma^2  I_p), \;
Y \mid X \sim \mathcal{N}\bigl(\beta_1^\top X + 2 \sin(\beta_2^\top X),  (\|X\|/p)^2\bigr),\, &&\textnormal{$\hat{\mu}$: from random forest regression},
\end{align*}
where we set the feature dimension $p = 20$ and $\sigma = 3$, and the parameters $\mu, \beta_1,\beta_2 \in \R^p$ are generated by drawing each element from the $\text{Unif}([0,1])$ distribution. Setting 1 reflects a scenario in which the estimator is constructed from a well-specified model and is therefore likely to be accurate, whereas Setting 2 represents a more general scenario in which the data arise from a more complex distribution, making accurate estimation more challenging. Recall that the proposed procedures are distribution-free and therefore always provide control of the FDR or the risk. However, as in standard conformal prediction, the quality of the estimators may affect the usefulness of the resulting outputs.

We first study the performance of the proposed procedure for constructing the uncertainty graph of the order with FDR control. We consider two variants of our proposed method and, for comparison, a p-value-based procedure:
\begin{enumerate}
    \item[(i)] Procedure~\ref{alg:fdr}: e-BH procedure + e-values $(e_{j_1,j_2})_{1 \leq j_1 < j_2 \leq m}$ from ~\eqref{eqn:e_val}.
    \item[(ii)] Procedure~\ref{alg:fdr} with boosting: U-eBH procedure + e-values $(e_{j_1,j_2})_{1 \leq j_1 < j_2 \leq m}$ from~\eqref{eqn:e_val}.
    \item[(iii)] p-value-based procedure: p-BH procedure + p-values $(p_{j_1,j_2})_{1 \leq j_1 < j_2 \leq m}$, described below.
\end{enumerate}
Here, we include the p-value-based method for comparison. While it does not come with a theoretical guarantee in our setting, it is well known that p-value-based procedures can perform reasonably well in practice across various settings, even when the corresponding theoretical guarantees do not apply. Specifically, 
for our problem, we construct the conformal p-value
and then apply the p-BH procedure to these p-values. The validity of the individual p-value $p_{j_1,j_2}$ in~\eqref{eqn:p_val} is established in Lemma~\ref{lem:p_val} in Section~\ref{sec:fwer}. However, due to the complex dependence structure of these p-values, it is not guaranteed that the p-BH procedure applied to these p-value controls the FDR control.

We generate a training dataset of size 500 and fit either a linear regression or a random forest regression to construct an estimator $\hat{\mu} : \X \rightarrow \R$. We then repeat the following steps 500 times: generate the calibration and test data, apply the procedures, and compute the number of true and false discoveries. Finally, we estimate the FDR and the power from the 500 trials. We use a calibration size of 200, and we repeat the simulations with test sizes 20, 50, and 100 at different levels $\alpha = 0.05,0.075,0.1,\ldots,0.2$.

\begin{figure}[htbp]
    \centering
    \includegraphics[width=0.8\linewidth]{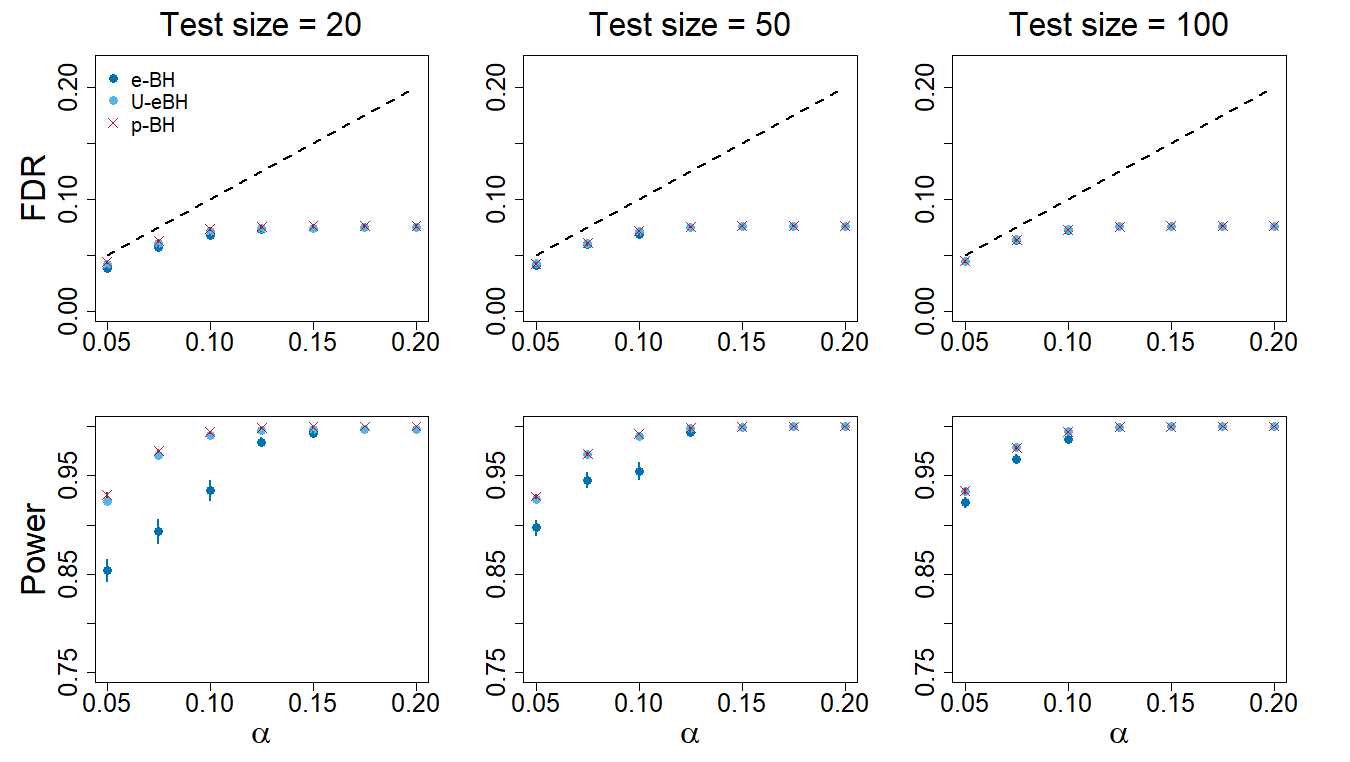}
    \caption{False discovery rate and power of Procedure~\ref{alg:fdr} (e-BH), its U-eBH variant (U-eBH), and the p-value-based method with~\eqref{eqn:p_val} (p-BH) under Setting 1, with calibration size 200 and test sizes 20, 50, and 100, with error bars. The dotted line corresponds to the line $y = x$.}
    \label{fig:fdr_linear}
\end{figure}

\begin{figure}[htbp]
    \centering
    \includegraphics[width=0.8\linewidth]{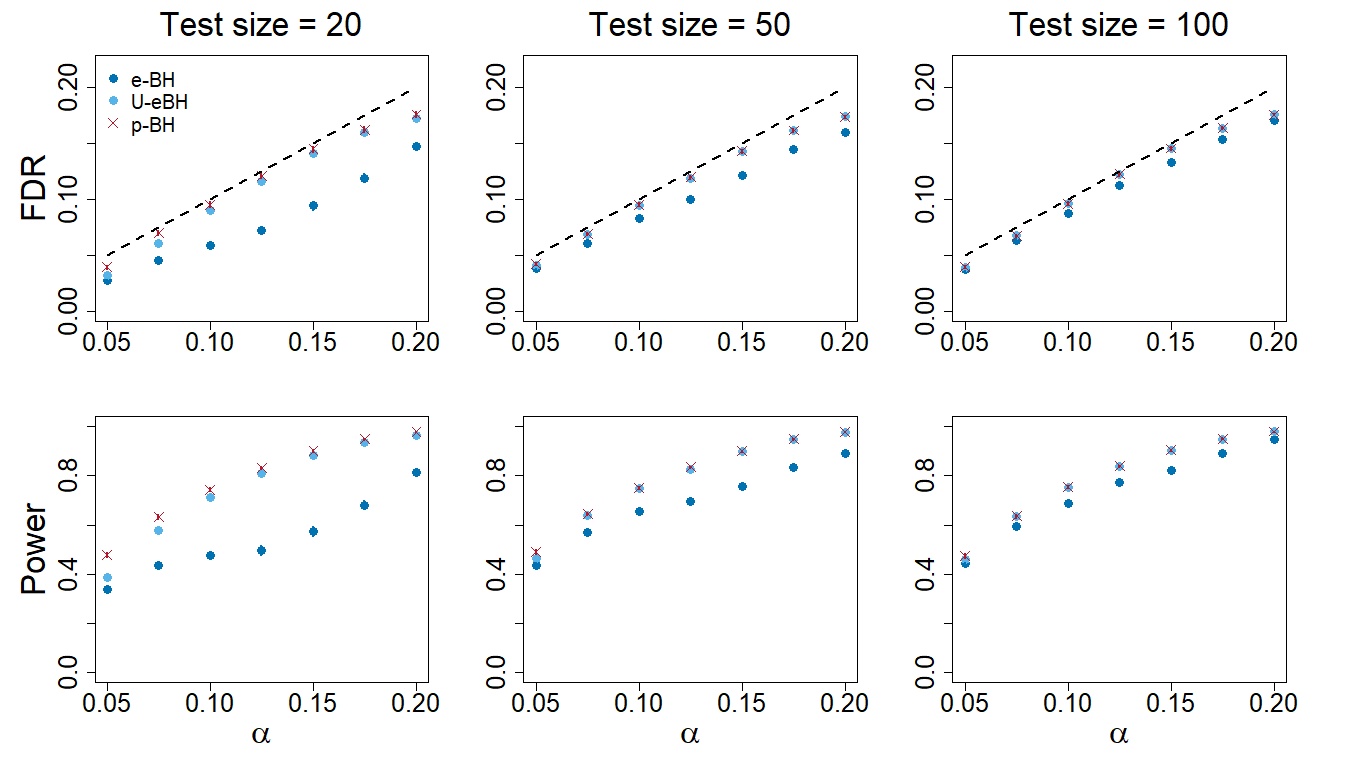}
    \caption{False discovery rate and power of Procedure~\ref{alg:fdr} (e-BH), its U-eBH variant (U-eBH), and the p-value-based method with~\eqref{eqn:p_val} (p-BH) under Setting 2, with calibration size 200 and test sizes 20, 50, and 100, with error bars. The dotted line corresponds to the line $y = x$.}
    \label{fig:fdr}
\end{figure}

The results for Settings 1 and 2 are shown in Figures~\ref{fig:fdr_linear} and~\ref{fig:fdr}, respectively. In both settings, the proposed procedures control the FDR across different target levels and test sizes, as guaranteed by our theoretical results, confirming that the FDR control holds regardless of the underlying data distribution or the quality of the estimator. The U-eBH variant of Procedure~\ref{alg:fdr} achieves some improvement in power over the base procedure, with performance nearly identical to that of the p-value-based method. Thus, our e-value-based procedures can be viewed as theoretically justified counterparts of the p-value-based procedure, retaining its favorable empirical performance while providing rigorous FDR control.

Note that the FDR is overall tightly controlled by the procedure, except in ``easy'' settings where high power can be achieved even with a very low FDR. On the other hand, the power of the procedure depends on the accuracy of the estimator. This coincides with the phenomenon observed in standard conformal prediction, where the prediction set becomes wider when the initial estimate is of low quality, in order to attain sufficient coverage. 

\begin{figure}[htbp]
    \centering
    \includegraphics[width=0.85\linewidth]{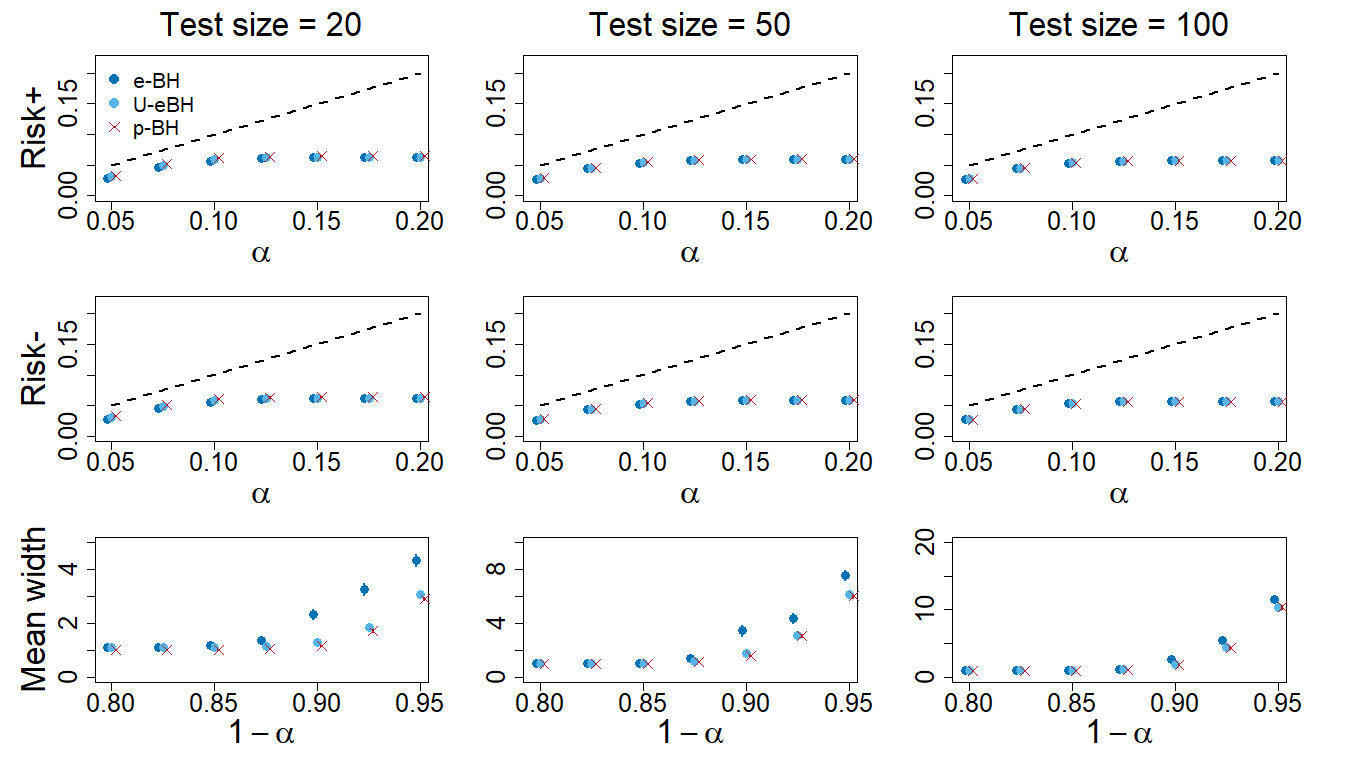}
    \caption{Risk and mean width of prediction intervals from Procedure~\ref{alg:rank_fdr} (e-BH), its U-eBH variant, and the p-value-based method (p-BH) under Setting 1, with calibration size 200 and test sizes 20, 50, and 100, with error bars. The dotted line corresponds to the line $y=x$.}
    \label{fig:sim_fdr_rank_linear}
\end{figure}

\begin{figure}[htbp]
    \centering
    \includegraphics[width=0.85\linewidth]{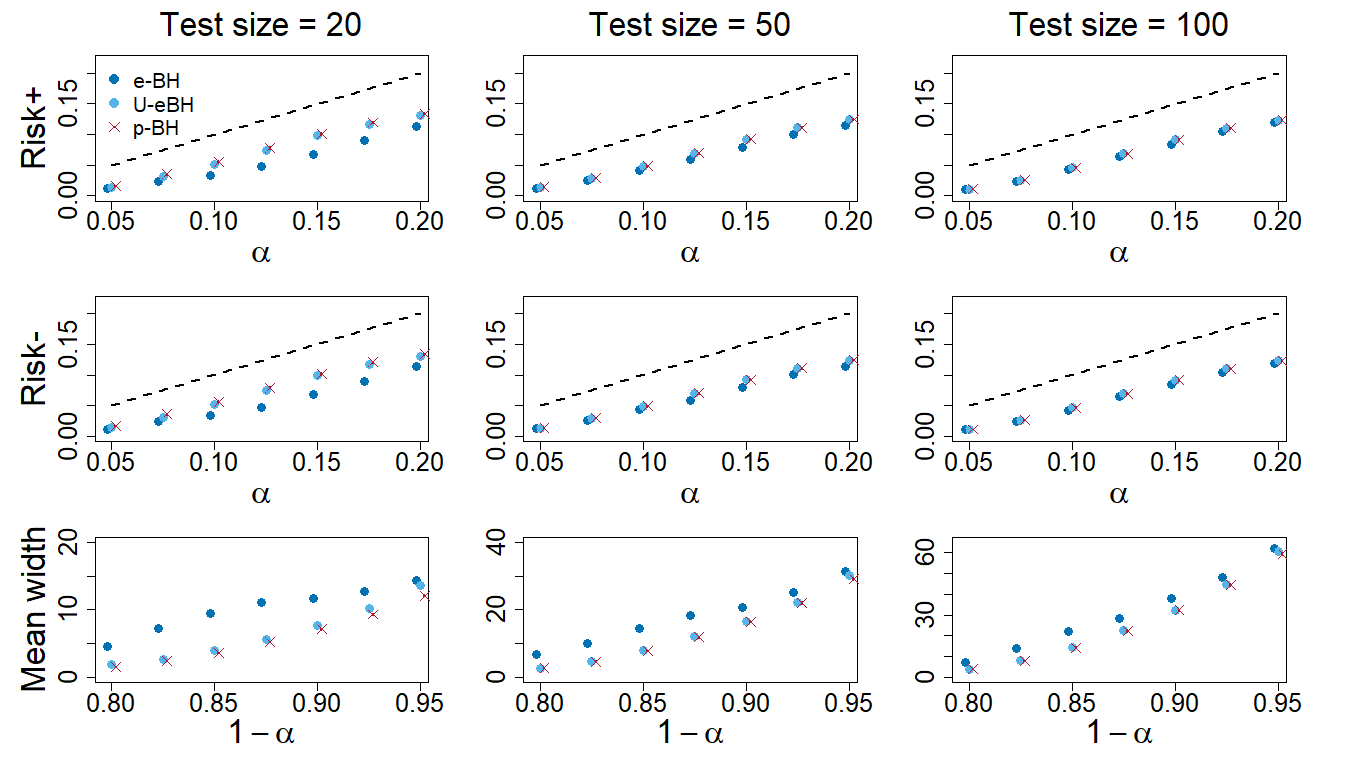}
    \caption{Risk and mean width of prediction intervals from Procedure~\ref{alg:rank_fdr} (e-BH), its U-eBH variant, and the p-value-based method (p-BH) under Setting 2, with calibration size 200 and test sizes 20, 50, and 100, with error bars. The dotted line corresponds to the line $y=x$.}
    \label{fig:sim_fdr_rank}
\end{figure}

Next, we investigate the procedure for simultaneous inference on the ranks, as described in Section~\ref{sec:rank}. Using the uncertainty graphs obtained from the three procedures in the previous experiment, we apply the steps of Algorithm~\ref{alg:rank_fdr} to construct prediction intervals for the ranks in each of the 500 trials. We then estimate \risk+ and \risk- in~\eqref{eqn:rank_risk}, along with the average coverage rate and average width of the prediction intervals. The results for the two settings are shown in Figures~\ref{fig:sim_fdr_rank_linear} and~\ref{fig:sim_fdr_rank}, respectively, illustrating that \risk+ and \risk- are controlled at the desired level by the proposed procedure. The average width of the prediction intervals is smaller in Setting 1, illustrating that, similarly to the power of the uncertainty graph, the width of the prediction intervals depends on the accuracy of the estimator---but, again, the validity of the procedure does not depend on the underlying data distribution or the quality of the estimator.

\subsection{Empirical data illustration}\label{sec:dti}

Here, we evaluate the proposed methods on a drug–target interaction (DTI) dataset. The data and pre-trained model are obtained from the DeepPurpose library \cite{huang2020deeppurpose}. The dataset contains 32,972 observations. Each observation consists of a drug–protein pair as covariates and a real-valued affinity score measuring the strength of interaction between the drug and the target protein as the response. We consider the task of predictive inference on the order and ranks of the test affinity scores, in the form of an uncertainty graph for the order and prediction intervals for the ranks. Inference on drug–target interactions is of broad importance in drug discovery; more specifically, inference on their relative ordering can help guide which drug candidates should be prioritized for development.

\begin{figure}[htbp]
    \centering
    \includegraphics[width=0.8\linewidth]{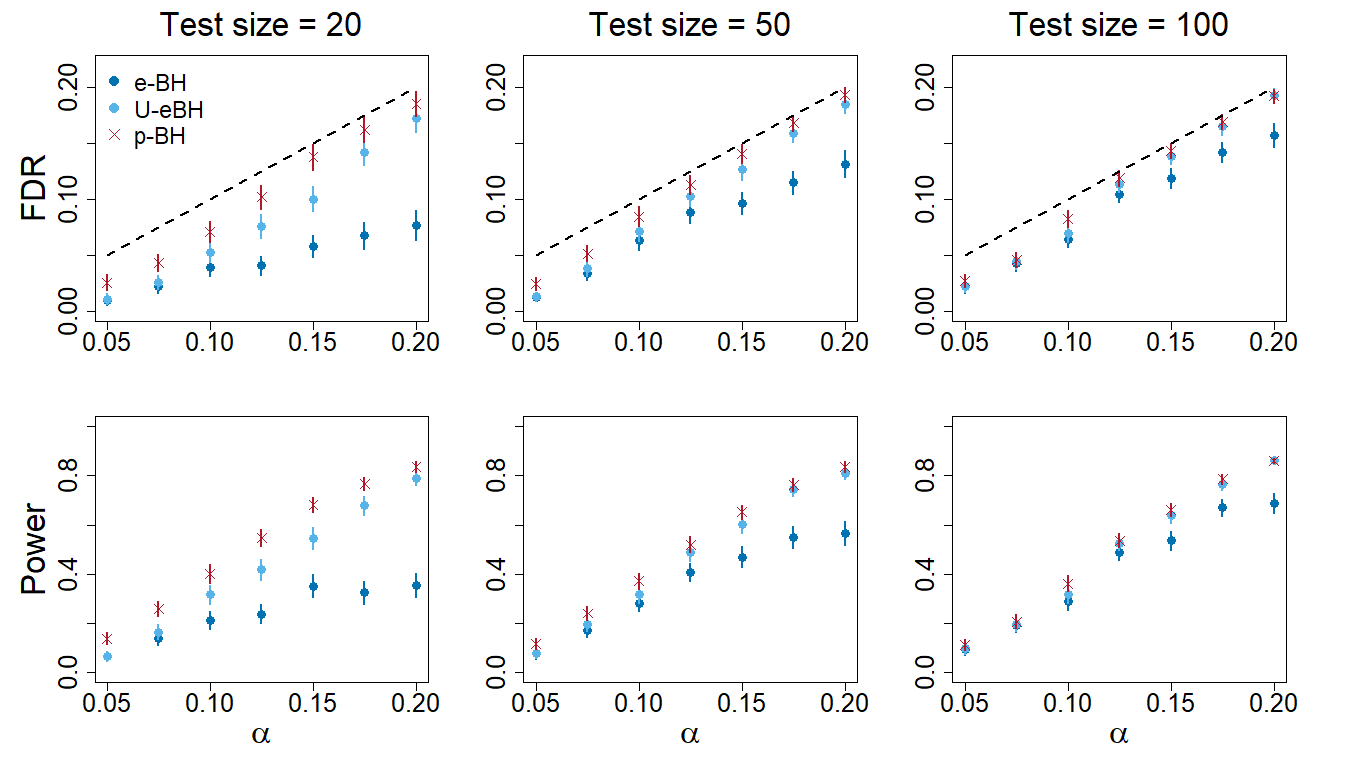}
    \caption{Results for the drug-target interaction dataset: False discovery rate and power of Procedure~\ref{alg:fdr}(e-BH), its U-eBH variant (U-eBH), and the p-value-based method (p-BH), under calibration size 200 and test sizes 20, 50, and 100, with error bars. The dotted line corresponds to the line $y = x$.}
    \label{fig:DTI_graph}
\end{figure}

For evaluation purposes, we construct 80 non-overlapping pairs of calibration and test sets, with a calibration size of $n = 200$, and repeat the experiment for three test sizes, $m = 20, 50,$ and $100$. First, for each pair, we apply Procedure~\ref{alg:fdr}, its U-eBH variant, and the p-value-based procedure at levels $\alpha = 0.05,0.075,0.1,\ldots,0.2$, to construct the uncertainty graph. We then estimate the FDR and the power based on the results from the 80 pairs. The results are shown in Figure~\ref{fig:DTI_graph}, illustrating that the FDR is tightly controlled at the desired level, across different levels and test sizes.

\begin{figure}[htbp]
    \centering
    \includegraphics[width=0.85\linewidth]{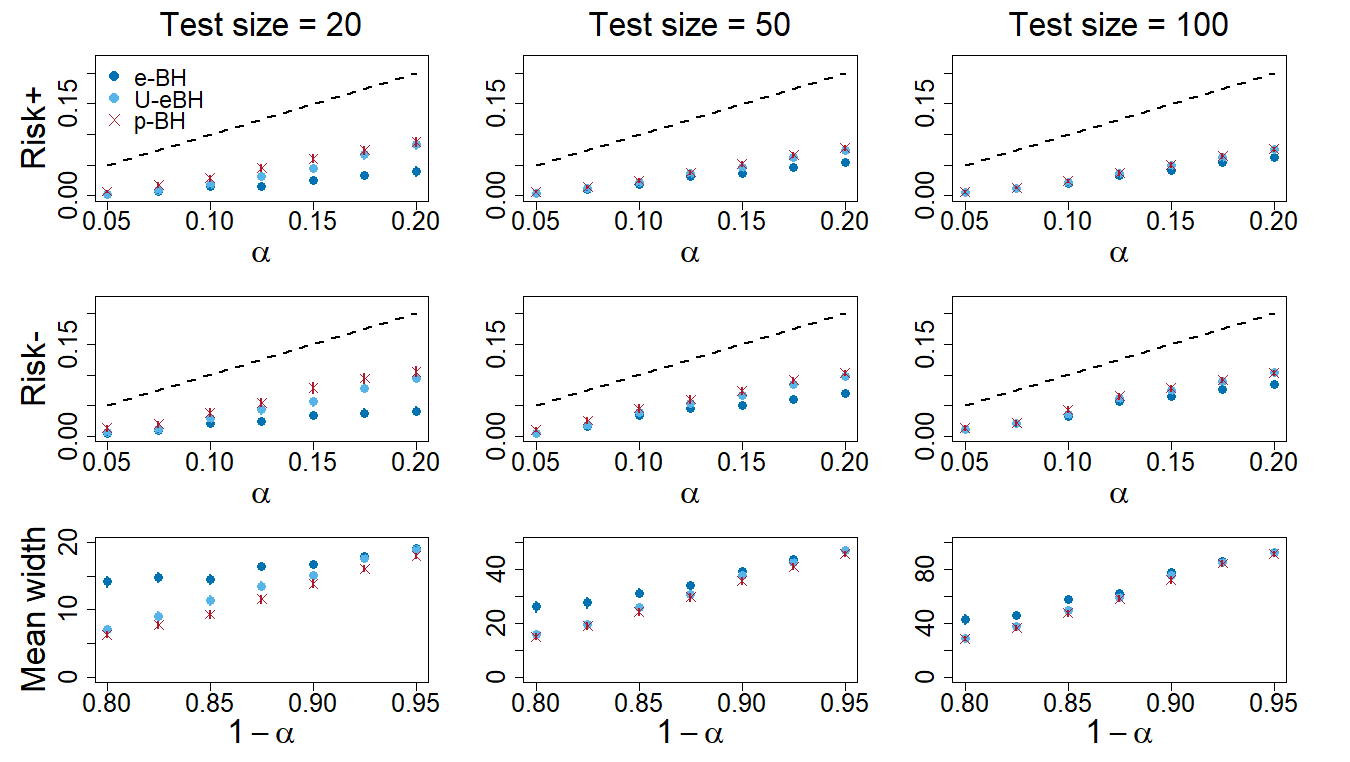}
    \caption{Results for the drug-target interaction dataset: Risk and the mean width of prediction intervals from Procedure~\ref{alg:rank_fdr} (e-BH), its U-eBH variant (U-eBH), and the p-value-based method (p-BH), under calibration size 200 and test sizes 20, 50, and 100, with error bars. The dotted line corresponds to the line $y = x$.}
    \label{fig:DTI_rank}
\end{figure}

We also construct simultaneous prediction intervals for the ranks using the uncertainty graphs obtained from the three procedures. The results are shown in Figure~\ref{fig:DTI_rank}, illustrating that the proposed procedure controls both $\risk+$ and $\risk-$ at the desired level.

\begin{figure}[htbp]
\centering
\resizebox{0.8\textwidth}{!}{%
\begin{tikzpicture}[x=1.4cm,y=1cm]

\draw[very thick] (0.4,0) -- (10.6,0);


\foreach \a/\b in {
    1/2,1/3,1/4,1/5,1/6,
    2/3,2/4,2/5,2/6,
    3/4,3/5,3/6,
    4/5,4/6,
    5/6,
    6/7,
    7/8,7/9,7/10,
    8/9,8/10,
    9/10
}{
    \pgfmathsetmacro{\d}{\b-\a}
    \pgfmathsetmacro{\h}{0.25+0.11*\d*\d}
    \draw[dotted,line width=1.2pt,red!70]
        (\a,0) .. controls
        (\a,\h) and (\b,\h)
        .. (\b,0);
}

\foreach \a/\b in {
    1/7,
    2/7,
    3/7,
    4/7,
    5/7,
    5/8,
    5/9,
    5/10,
    6/8
}{
    \pgfmathsetmacro{\d}{\b-\a}
    \pgfmathsetmacro{\h}{0.25+0.11*\d*\d}
    \draw[solid,line width=0.5pt,blue!20!gray!70]
        (\a,0) .. controls
        (\a,\h) and (\b,\h)
        .. (\b,0);
}

\foreach \x in {1,...,10} {
    \fill[black] (\x,0) circle (2pt);
}

\foreach \x/\drug/\target in {
    1/10184653/PLK4,
    2/11314340/LZK,
    3/153999/NIM1,
    4/9915743/MEK4,
    5/153999/ABL1(H396P)p,
    6/10113978/BMPR1B,
    7/11427553/FYN,
    8/5494449/ARK5,
    9/16038120/CDK7,
    10/5329102/AAK1
}{
    \node[rotate=45,anchor=north east,align=right,
          font=\scriptsize] at (\x,-0.12)
          {\drug\\\target};
}

\end{tikzpicture}%
}

\caption{Illustration of the uncertainty graph obtained from the drug--target interaction dataset, with a test set of size $10$. For clarity, only the shortest solid edge to the left and right of each test point is displayed.}
\label{fig:dti_graph_ex}

\end{figure}
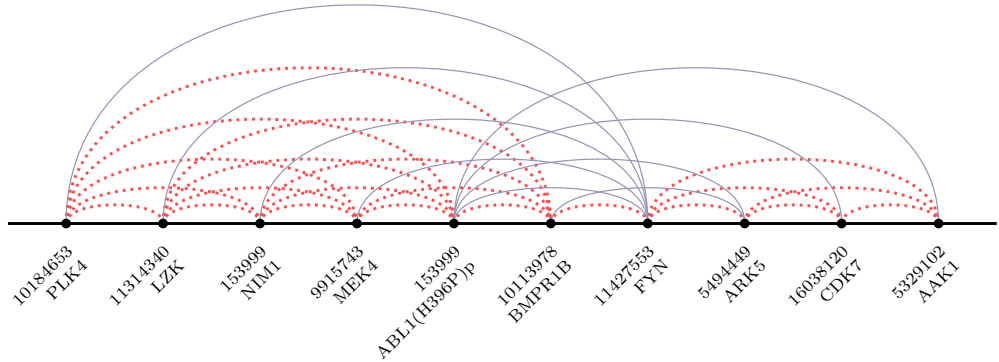

In addition to the above evaluation of the procedure on the DTI dataset, we provide a visual illustration of what the output of the procedure looks like in practice. Specifically, Figure~\ref{fig:dti_graph_ex} shows the uncertainty graph constructed for 10 test points using 200 calibration samples. The drug--target pairs are arranged along the horizontal line in increasing order of their predicted outcomes. Solid edges indicate pairwise ordering relations detected with certainty by the proposed procedure (Algorithm~\ref{alg:fdr}), while dotted edges indicate pairs for which the ordering remains uncertain. For simplicity, some of the solid edges that follow from other solid edges through consistency are omitted.

\section{Discussion}

In this work, we developed a distribution-free methodology for predictive inference on the ordering of multiple unobserved outcomes. We proposed a multiple-testing-inspired approach that yields a consistent uncertainty graph with FDR control and showed that it also yields simultaneous prediction intervals for the ranks with risk control. Empirical results demonstrate that the proposed methods controls the error rates at the target levels while attaining reasonable power and interval widths.

Many open questions remain. Conformal prediction and conformal selection have been studied and extended in various settings beyond exchangeable data and standard marginal guarantees. Similar extensions may be explored for inference on ordering, offering opportunities to generalize the methods developed in this work. We leave these directions for future work.

\subsection*{Acknowledgment}
Z.R.~is supported by NSF grant DMS-2413135 and Wharton Analytics. Y.L.~is supported in part by 
NIH R01-AG065276, R01-GM139926, NSF 2210662, P01-AG041710, R01-CA222147, 
ARO W911NF-23-1-0296,
NSF 2046874, ONR N00014-21-1-2843, and the Sloan Foundation.

\bibliographystyle{plainnat}
\bibliography{bib}

\newpage

\appendix


\section{Additional experimental results}

\subsection{Stability of the U-eBH boosting}\label{sec:exp_uebh}

Here, we empirically investigate the stability of the U-eBH variant of the proposed procedure. Recall that the U-eBH procedure draws $U \sim \mathrm{Unif}[0,1]$ and divides all e-values by $U$ before applying the e-BH procedure. We examine whether the results vary substantially depending on the realized value of $U$.

We repeat the simulations in Section~\ref{sec:sim} with test size $m=20$, but in each trial, we generate 100 independent realizations of $U$ and apply the U-eBH-based procedure for each realization. We then compute the standard deviations of the false discovery proportion (FDP) and empirical power across the 100 realizations. We repeat this experiment over 500 trials and report the average standard deviations in Table~\ref{tab:stability}. The results show that the variability in the FDP and power across different realizations of $U$ is small in both settings.

\begin{table}[ht]
\centering
\begin{tabular}{llrrrrrrr}
  \hline
 & $\alpha$ & 0.05 & 0.075 & 0.1 & 0.125 & 0.15 & 0.175 & 0.2 \\
  \hline
  \hline
\multirow{2}{*}{Setting 1}
  & FDR & 0.0004 & 0.0004 & 0.0003 & 0.0001 & 0.0000 & 0.0000 & 0.0000 \\ 
  & Power & 0.0116 & 0.0097 & 0.0060 & 0.0010 & 0.0002 & 0.0000 & 0.0000 \\ 
  \hline
\multirow{2}{*}{Setting 2}
  & FDR & 0.0009 & 0.0028 & 0.0056 & 0.0064 & 0.0070 & 0.0053 & 0.0048 \\
  & Power & 0.0099 & 0.0246 & 0.0414 & 0.0440 & 0.0467 & 0.0343 & 0.0283 \\
  \hline
\end{tabular}
\caption{Mean standard deviations of the FDP and empirical power for the U-eBH variant of Procedure~\ref{alg:fdr} in Settings 1 and 2, with a test size of 20.}
\label{tab:stability}
\end{table}

\subsection{Conformal ordering with FWER control}\label{sec:fwer_exp}

Here, we provide simulation results for the methods discussed in Section~\ref{sec:fwer}. We investigate the performance of the Procedures~\ref{alg:fwer}  and~\ref{alg:fwer_par} which provides an uncertainty graph with FWER control. For data generating process, we follow Setting 2 in Section~\ref{sec:sim}. We generate training data of size 500 and fit a random forest regression to construct an estimator $\hat{\mu} : \X \rightarrow \R$. We repeat the following steps 500 times: generate the calibration and test data, apply Procedures~\ref{alg:fwer} and~\ref{alg:fwer_par} and construct the corresponding rank prediction intervals, and compute the numbers of true and false discoveries. We then estimate the FWER and power of the uncertainty graph, as well as the coverage rate and average width of the rank prediction intervals, across the 500 trials. We use a calibration size of 1000, and we repeat the simulations with test sizes 5, 10, 20, and 50 at different levels $\alpha = 0.05,0.075,0.1,\ldots,0.2$. 

\begin{figure}[htbp]
    \centering
    \includegraphics[width=0.9\linewidth]{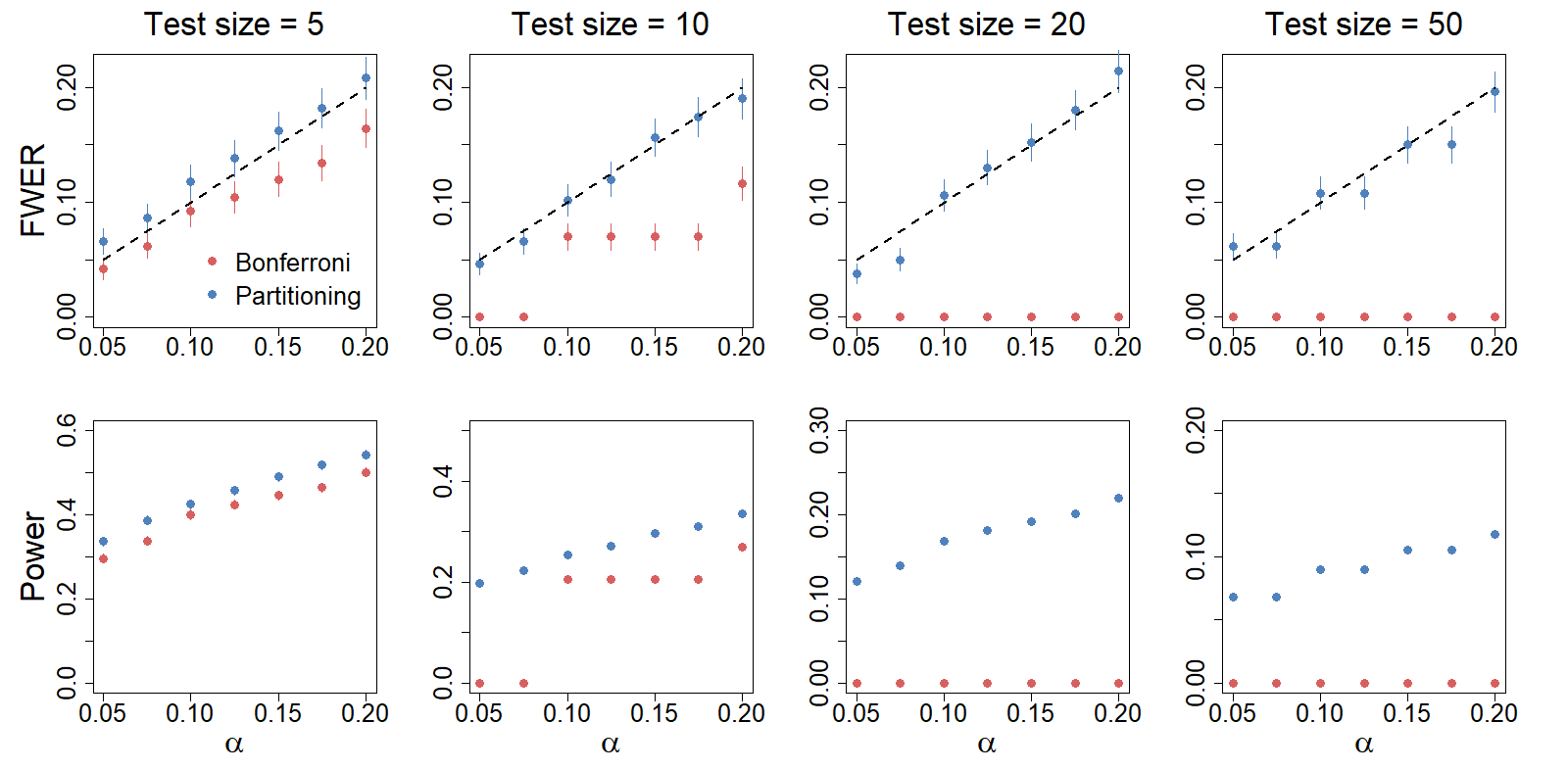}
    \caption{Family-wise error rate and power of Procedures~\ref{alg:fwer} and~\ref{alg:fwer_par}, under calibration size 1000 and test sizes 5, 10, 20, and 50. The dotted line corresponds to the line $y = x$.}
    \label{fig:fwer}
\end{figure}

\begin{figure}[htbp]
    \centering
    \includegraphics[width=0.9\linewidth]{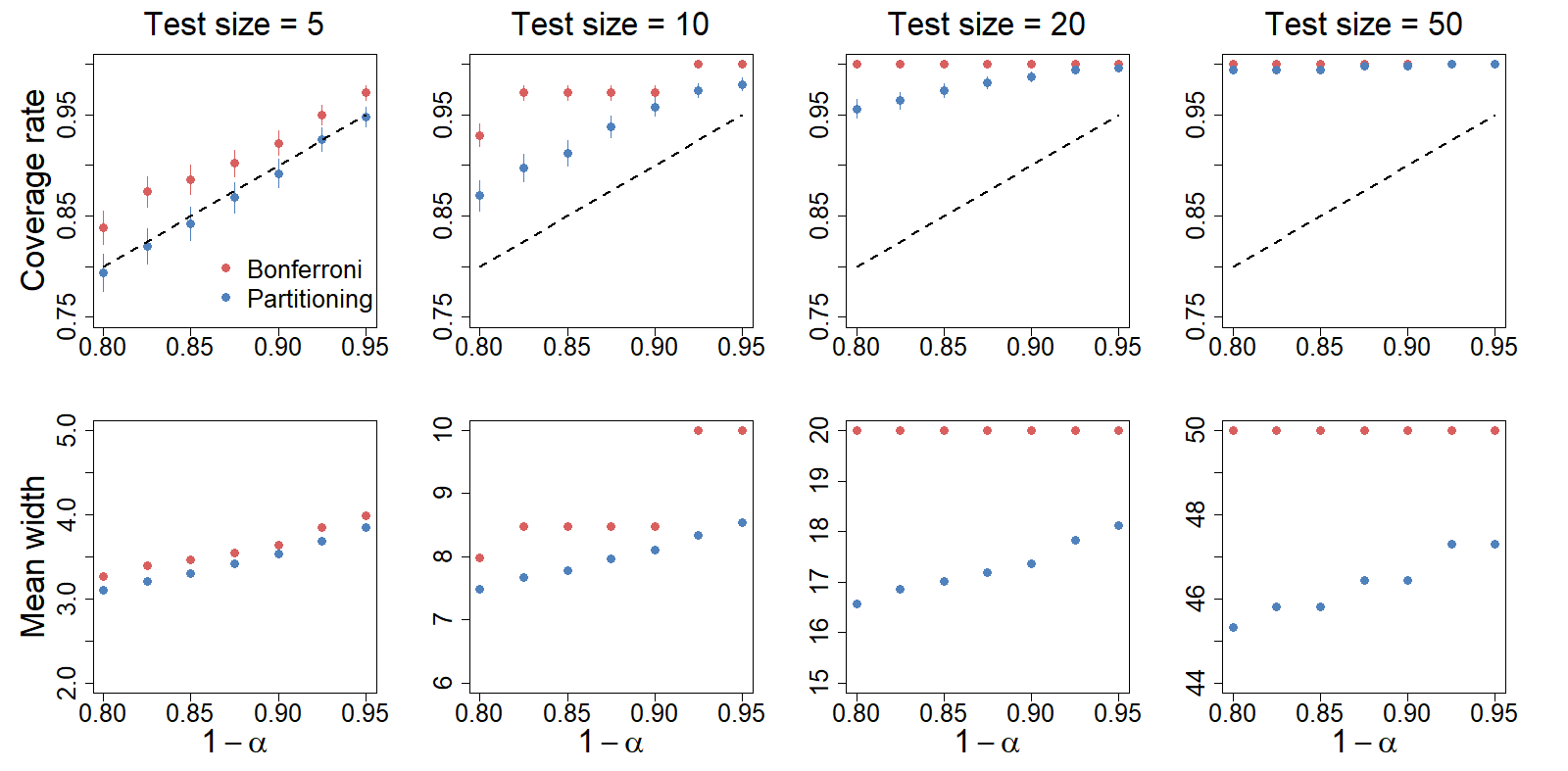}
    \caption{Coverage rate and Average width of the rank prediction intervals induced by Procedures~\ref{alg:fwer} and~\ref{alg:fwer_par}, under calibration size 1000 and test sizes 5, 10, 20, and 50. The dotted line corresponds to the line $y = x$.}
    \label{fig:fwer_rank}
\end{figure}

The results are shown in Figure~\ref{fig:fwer} and~\ref{fig:fwer_rank}. Both procedures control the FWER across different settings, but Procedure~\ref{alg:fwer} becomes increasingly conservative as the test size increases. In contrast, Procedure~\ref{alg:fwer_par} generally achieves tighter FWER control while attaining higher power. A similar pattern is observed for the induced rank prediction intervals. Both methods achieve coverage rates above the nominal level $1-\alpha$, but the partitioning-based method~\ref{alg:fwer_par} generally yields coverage rates closer to the nominal level and shorter intervals.

\begin{figure}[htbp]
    \centering
    \includegraphics[width=0.9\linewidth]{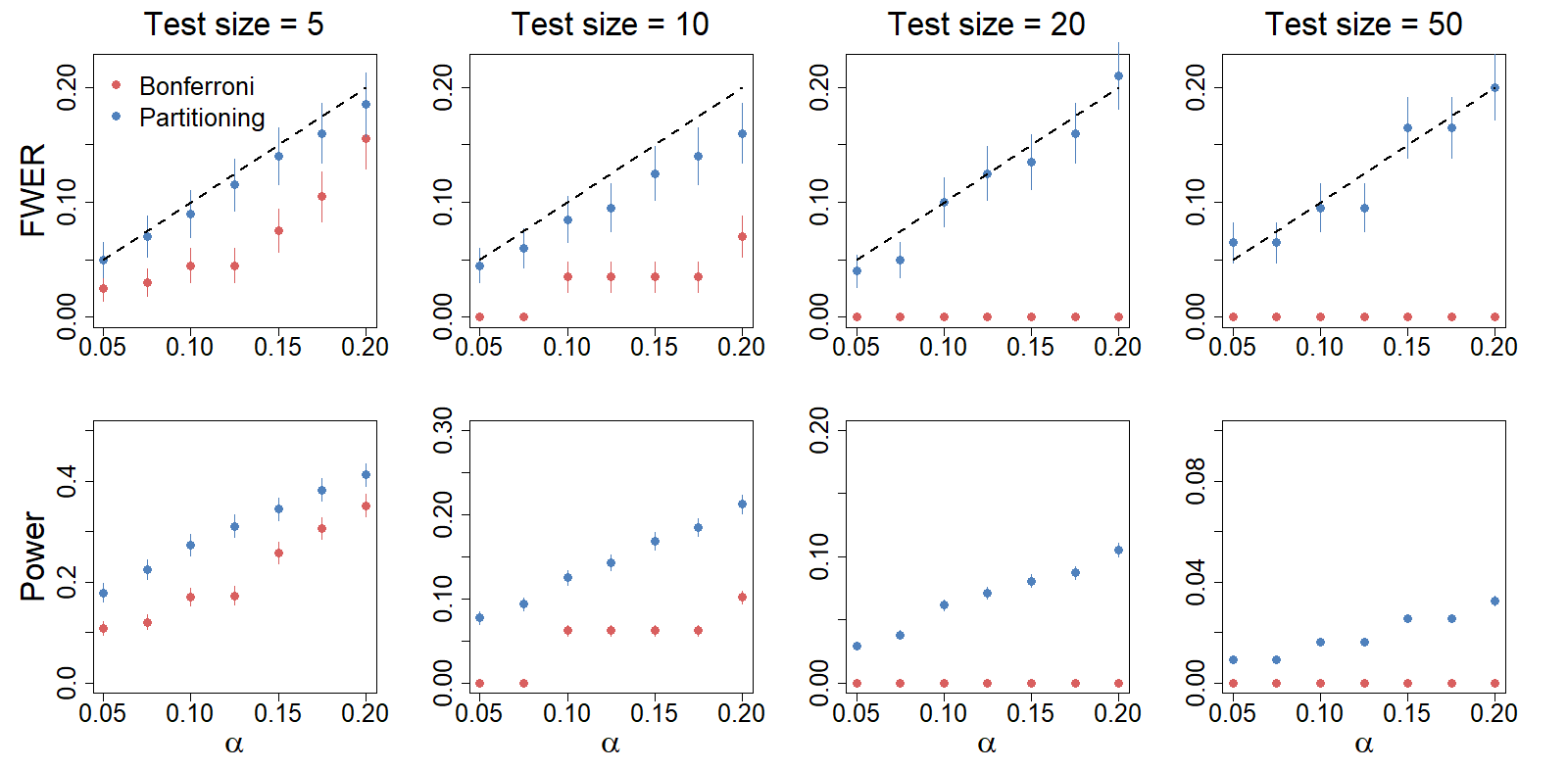}
    \caption{Results for the Drug-target interaction dataset: Family-wise error rate and power of Procedures~\ref{alg:fwer} and~\ref{alg:fwer_par}, under calibration size 1000 and test sizes 5, 10, and 20. The dotted line corresponds to the line $y = x$.}
    \label{fig:dti_fwer}
\end{figure}

\begin{figure}[htbp]
    \centering
    \includegraphics[width=0.9\linewidth]{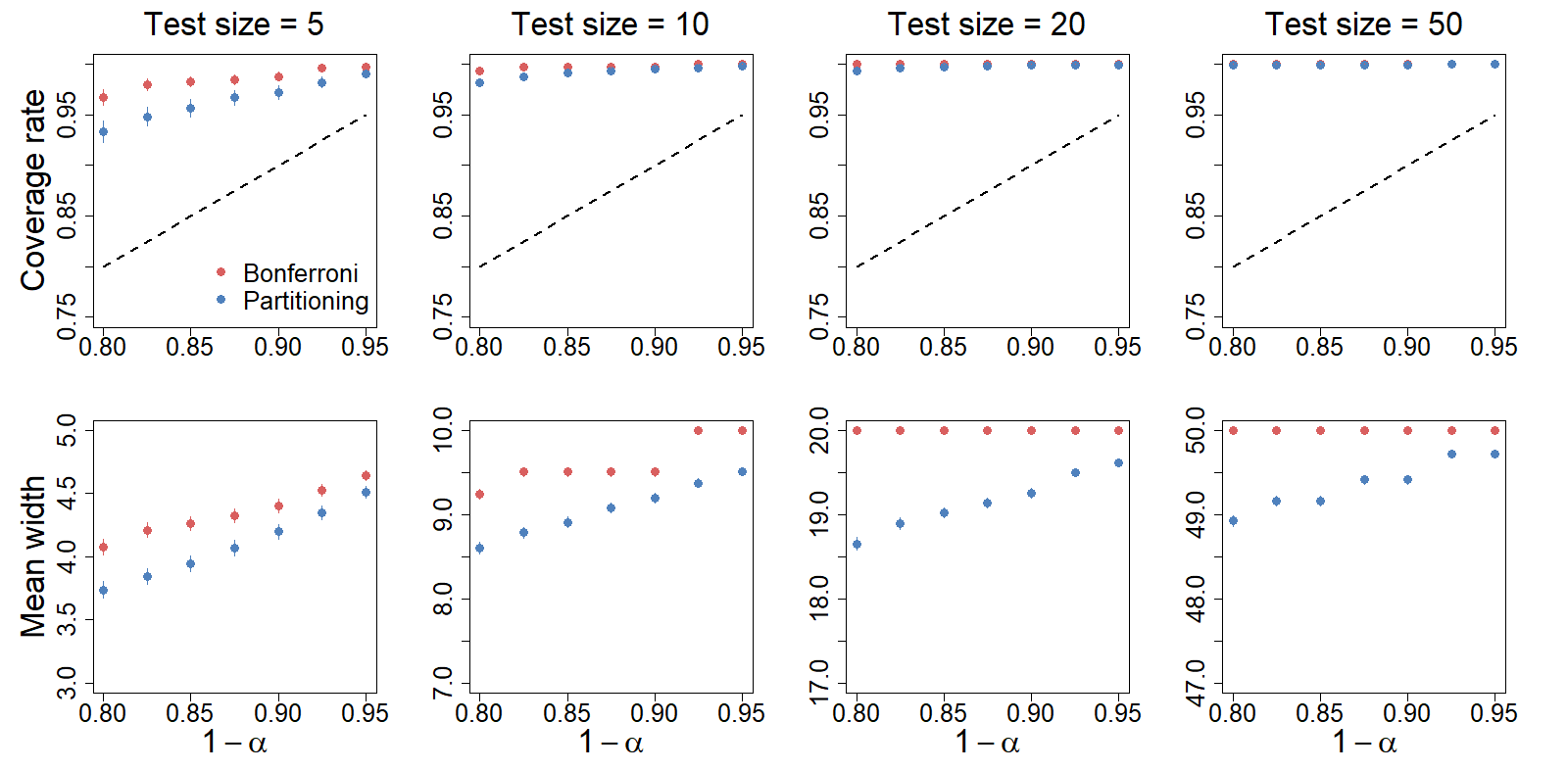}
    \caption{Results for the Drug-target interaction dataset: Coverage rate and mean width of the rank prediction intervals induced by Procedures~\ref{alg:fwer} and~\ref{alg:fwer_par}, under calibration size 1000 and test sizes 5, 10, and 20. The dotted line corresponds to the line $y = x$.}
    \label{fig:dti_fwer_rank}
\end{figure}

We also provide results for the drug–target interaction dataset in Figure~\ref{fig:dti_fwer} and~\ref{fig:dti_fwer_rank}. The steps overall follow those in Section~\ref{sec:dti}, with the main algorithm replaced by Procedure~\ref{alg:fwer} and~\ref{alg:fwer_par}. The results follow a similar pattern to those observed in the simulations. As discussed earlier, the FWER-based procedures tend to be conservative, and we generally recommend the FDR-based procedures unless the test size is substantially smaller than the calibration size.

\section{Technical proofs}

\subsection{Proof of Lemma~\ref{lem:p_val}}
\label{app:proof-pval}

By the exchangeability of the pairs $((X_{2i-1},Y_{2i-1}), (X_{2i},Y_{2i}))_{i \in [n_1]}$ and $((X_{n+j_1},Y_{n+j_1}),(X_{n+j_2},Y_{n+j_2}))$, we have that for any score function $s:(\X \times \R)^2 \rightarrow \R$, the conformal p-value
\[\tilde{p}_{j_i,j_2} := \frac{\sum_{i=1}^{n_1} \One{s((X_{2i-1},Y_{2i-1}), (X_{2i},Y_{2i})) \ge s((X_{n+j_1},Y_{n+j_1}),(X_{n+j_2},Y_{n+j_2}))} + 1}{n_1+1}\]
is super-uniform~\citep{vovk2005algorithmic}. Now define
\begin{multline*}
    s((x_1,y_1), (x_2,y_2)) = |\mh(x_1) - \mh(x_2)|\cdot \One{(y_1 - y_2)(\mh(x_1) - \mh(x_2)) \le 0}\\
    + (-\infty) \cdot \One{(y_1 - y_2)(\mh(x_1) - \mh(x_2)) > 0}.
\end{multline*}
Then on the event $\{(Y_{n+j_1} - Y_{n+j_2})\cdot(\mh(X_{n+j_1}) - \mh(X_{n+j_2})) \le 0\}$,
\begin{multline*}
    s((X_{2i-1},Y_{2i-1}), (X_{2i},Y_{2i})) \ge s((X_{n+j_1},Y_{n+j_1}), (X_{n+j_2},Y_{n+j_2}))\\
    \iff |\mh(X_{2i-1}) - \mh(X_{2i})| \ge |\mh(X_{n+j_1}) - \mh(X_{n+j_2})| \text{ and } (Y_{2i-1} - Y_{2i})(\mh(X_{2i-1}) - \mh(X_{2i})) \le 0.
\end{multline*}
As a result, for any $\alpha \in [0,1]$,
\begin{align*}
\PP{p_{j_1,j_2} \le \alpha , H_{j_1,j_2} \text{ holds}} 
= \PP{\tilde{p}_{j_1,j_2} \le \alpha, H_{j_1,j_2} \text{ holds}} \le \alpha.
\end{align*}
This completes the proof.

\subsection{Proof of Theorem~\ref{thm:fwer}}
\label{app:proof-fwer}
By lemma~\ref{lem:p_val}, we have
\[\PP{p_{j_1,j_2} \leq \alpha \textnormal{ and } H_{j_1,j_2} \textnormal{ holds} } \leq \alpha\]
for all $1 \leq j_1 < j_2 \leq m$, for any $\alpha \in (0,1)$. Therefore,
\begin{align*}
    &\FWER = \PP{\exists\, 1 \leq j_1 < j_2 \leq m \text{ such that } (j_1,j_2) \in \mathcal{R}, H_{j_1,j_2} \text{ holds}}\\
    &\leq \sum_{1 \leq j_1 < j_2 \leq m} \PP{(j_1,j_2) \in \mathcal{R}, H_{j_1,j_2} \text{ holds}} = \sum_{1 \leq j_1 < j_2 \leq m} \PP{p_{j_1,j_2} \leq 2\alpha/m(m-1), H_{j_1,j_2} \text{ holds}}\\
    &\leq \sum_{1 \leq j_1 < j_2 \leq m} 2\alpha/m(m-1) = \alpha,
\end{align*}
as desired.

Consistency follows directly from the definition of the p-value in~\eqref{eqn:p_val}, since $p_{j_1,j_2}$ is a monotone decreasing function of $|\mh(X_{n+j_1}) - \mh(X_{n+j_2})|$.

\subsection{Proof of Theorem~\ref{thm:fwer_par}}
\label{app:proof-fwer-par}
Observe that $\FWER$ can be expressed as
\begin{align*}
    \FWER = \PP{\max_{\substack{1 \leq j_1 < j_2 \leq m}} |\hat{\mu}(X_{n+j_1})-\hat{\mu}(X_{n+j_2})| \cdot \One{(\hat{\mu}(X_{n+j_1})-\hat{\mu}(X_{n+j_2}))(Y_{n+j_1}-Y_{n+j_2}) \leq 0} > T},
\end{align*}
since the event inside the probability is equivalent to the existence of at least one pair $(j_1,j_2)$ satisfying both $(\hat{\mu}(X_{n+j_1})-\hat{\mu}(X_{n+j_2}))(Y_{n+j_1}-Y_{n+j_2}) \leq 0$ and $|\hat{\mu}(X_{n+j_1})-\hat{\mu}(X_{n+j_2})| > T$---since $T \geq 0$ by construction. The maximum above is exchangeable with the group-wise scores $S_{G_1}, \ldots, S_{G_K}$ by construction. Thus, the standard conformal prediction argument immediately implies that $\FWER \leq \alpha$.

Consistency follows directly from the definition of the rejection set using a common threshold $T$.

\subsection{Proof of Theorem~\ref{thm:hardness}}
\label{app:proof-hardness}
For notational simplicity, let $Z_i = (X_i,Y_i)$ for $i \in [n+m]$.
Let $I \in \mathcal{S}_{n+m}$ denote the order of the $n+m$ samples according 
to their predicted values $f(X_i)$ in ascending order, that is, 
\begin{align*}
    f(X_{I_1}) \le f(X_{I_2}) \le \ldots \le f(X_{I_{n+m}}).
\end{align*}
Define 
\begin{align*}
    L = \min\big\{i \in [n+m] : Z_{I_i} \text{ is a test sample}\big\}.
\end{align*}
to be the index of the first test sample in the ordered sequence. For each fixed $r\in[n+m]$, define the modified responses
\[
\tilde Y_i =
\begin{cases}
\max_{\ell\in[n+m]} Y_\ell + 1, & i=I_r,\\
Y_i, & i\neq I_r,
\end{cases}
\qquad i\in[n+m],
\]
and let $Q_r$ denote the joint distribution of
$(\tilde Z_i)_{i \in [n+m]}$, where $\tilde{Z}_i = (X_i,\tilde Y_i)$. Since the modification is permutation-equivariant, $Q_r$ is exchangeable.

Since the procedure controls the FWER also under the exchangeable distribution $Q_r$, we have
\begin{align*}
\mathbb{P}\big\{\exists (j_1,j_2) \in \widetilde{\mathcal{R}}, (f(X_{n+j_1}) - f(X_{n+j_2}))\cdot(\tilde Y_{n+j_1} - \tilde Y_{n+j_2}) \le 0 \big\} \leq \alpha,
\end{align*}
where $\widetilde{\mathcal{R}}$ denotes the rejection set produced by the procedure when applied to the modified data $(\tilde Z_i)_{i\in [n+m]}$.

On the event $\{\mathcal{R}\neq \varnothing, L=r\}$,
the modified unit is a test one, and therefore the input to 
the procedure is the same for the original data $(Z_i)_{i \in [n+m]}$
and the modified data $(\tilde Z_i)_{i\in [n+m]}$, implying $R = \widetilde{\mathcal{R}}$.
By consistency, the pair involving $Z_{I_L}$ and the test sample with the largest 
predicted value---indexed by $I_R$---must be rejected. 
That is, we certify that $Y_{I_L} < Y_{I_R}$.
Meanwhile, by the construction of $\tilde Y_i$, we have $\tilde Y_{I_L} > \tilde Y_{I_R}$. 
Therefore, 
\[
(I_L \wedge I_R - n, I_L \vee I_R - n) \in \mathcal{R} = \widetilde{\mathcal{R}} \text{ and } 
(f(X_{I_L}) - f(X_{I_R}))\cdot(\tilde Y_{I_L} - \tilde Y_{I_R}) \le 0.
\]
Clearly, this event is contained in the event considered in the FWER control above, implying that
\begin{align*}
\PP{\mathcal{R} \neq \varnothing, L = r} \leq \min\{\alpha,p_r\}, \quad \text{ where } p_r = \PP{L = r} = \frac{\binom{n+m-r}{m-1}}{\binom{n+m}{m}}.
\end{align*} 
Therefore,
\begin{multline*}
    \PP{\mathcal{R}=\varnothing}=1-\PP{\mathcal{R}\neq\varnothing} = 1 - \sum_{r=1}^{n+1}\PP{\mathcal{R} \neq \varnothing, L = r} \geq 1-\sum_{r=1}^{n+1}\min\{\alpha,p_r\}\\
    =\sum_{r=1}^{n+1} p_r - \sum_{r=1}^{n+1}\min\{\alpha,p_r\} = \sum_{r=1}^{n+1}(p_r-\alpha)_+.
\end{multline*}

\subsection{Proof of Theorem~\ref{thm:rank_interval}}
\label{app:proof-rank-interval}
Define $E_k$ as the set of edges in $\mathcal{R}_{\rm ord}$ with left node $k$, i.e.,
\begin{equation*}
    E_k = \{(k,l) : l > k, (k,l) \in \mathcal{R}_{\rm ord}\}.
\end{equation*}
Now observe that if all pairs in $E_k$ are true discoveries, 
then there are at least $|E_k|$ values among the $Y_{n+l}$'s that are larger than $Y_{n+\hat I_k}$, which implies that the rank of $Y_{n+\hat I_k}$ is at most $m-|E_k|$. Therefore, if $R_{\hat I_k} > m-|E_k| = \hat{u}_k$, then $E_k$ must include a false discovery. Hence, we have
\begin{align*}
    &\PP{\exists\, j \in [m] \textnormal{ such that } R_j \notin \hat{C}_j} = \PP{\exists\, j \in [m] \textnormal{ such that } R_{\hat I_j} \notin \tilde{C}_{j}}\\
    &= \PP{\exists\, j \in [m] \textnormal{ such that } R_{\hat I_j} > \hat{u}_{j}} \leq \PP{\exists\, j \in [m] \textnormal{ such that $E_{j}$ includes a false discovery}}\\
    &= \PP{\mathcal{R} \textnormal{ includes a false discovery}} \leq \alpha,
\end{align*}
where the last inequality holds by the assumption that $\mathcal{R}$ controls the FWER. It follows that
\[
\PP{\forall\, j \in [m], R_j \in \hat{C}_j} \geq 1-\alpha,
\]
as desired.

\subsection{Proof of Corollary~\ref{cor:rank_fwer_two_sided}}
By arguments similar to those in the proof of Theorem~\ref{thm:rank_interval}, we have
\[\PP{\forall\, j \in [m], R_j \leq \hat{u}_{\hat{I}_j}} \geq 1 - \alpha/2, \quad \text{and} \quad \PP{\forall\, j \in [m], R_j \geq \hat{l}_{\hat{I}_j}} \geq 1 - \alpha/2.\]
Applying the union bound completes the proof.

\subsection{Proof of Lemma~\ref{lem:fdr}}
\label{app:proof-eval}
Let us write $A_{i_1,i_2} = \One{(Y_{i_1}-Y_{i_2})(\hat{\mu}(X_{i_1}) - \hat{\mu}(X_{i_2})) \le 0}$ 
for $1 \leq i_1,i_2 \leq n+m$. For each $1 \leq j_1 < j_2 \leq m$, define a perturbed version of the threshold $T_{j_1,j_2}$ as
\begin{align*}
    &\tilde{T}_{j_1,j_2} = \inf\Big\{t>0 : \widetilde{\FDP}_{j_1,j_2}(t) \leq \alpha\Big\}, \text{ where }\\
    &\widetilde{\FDP}_{j_1,j_2}(t) = \frac{\sum_{i=1}^{n_1} \One{|\hat{\mu}(X_{2i-1}) - \hat{\mu}(X_{2i})| > t}\cdot A_{2i-1,2i} + \One{|\hat{\mu}(X_{n+j_1}) - \hat{\mu}(X_{n+j_2})| > t}\cdot A_{n+j_1,n+j_2}}{1 \vee \sum_{\substack{1\le j_1'<j_2'\le m\\
j_1',j_2'\notin\{j_1,j_2\}}} \One{|\hat{\mu}(X_{n+j_1'}) - \hat{\mu}(X_{n+j_2'})| > t}}\\
    &\hspace{130mm}\cdot \frac{(m-2)(m-3)/2}{n_1+1}.
\end{align*}
Of course, $\tilde T_{j_1,j_2}$ is not computable since it depends on the unknown $A_{n+j_1,n+j_2}$, but we will 
show that on the event $\{A_{n+j_1,n+j_2} = 1, |\hat{\mu}(X_{n+j_1}) - \hat{\mu}(X_{n+j_2})| > T_{j_1,j_2}\}$, 
we have $\tilde{T}_{j_1,j_2} = T_{j_1,j_2}$, which will allow us to use the exchangeability condition. 

Observe that $\widetilde{\FDP}_{j_1,j_2}(t) \leq \widehat{\FDP}_{j_1,j_2}(t)$ holds for any $t$ by construction. It follows that $\tilde{T}_{j_1,j_2} \leq T_{j_1,j_2}$ for any $1 \leq j_1 < j_2 \leq m$. Next, suppose that $A_{n+j_1,n+j_2} = 1$ and $|\hat{\mu}(X_{n+j_1}) - \hat{\mu}(X_{n+j_2})| > T_{j_1,j_2}$ hold. This immediately implies $|\hat{\mu}(X_{n+j_1}) - \hat{\mu}(X_{n+j_2})| > \tilde{T}_{j_1,j_2}$, and thus for any $\eps > 0$, there exists $\tilde{T}_{j_1,j_2} < t_0 < \min\{\tilde{T}_{j_1,j_2}+\eps, |\hat{\mu}(X_{n+j_1}) - \hat{\mu}(X_{n+j_2})|\}$ such that $\widetilde{\FDP}_{j_1,j_2}(t_0) \leq \alpha$ holds. Note that by the construction of $t_0$ and the assumption of $A_{n+j_1,n+j_2} = 1$, we have $\widehat{\FDP}_{j_1,j_2}(t_0) = \widetilde{\FDP}_{j_1,j_2}(t_0) \leq \alpha$, which implies $T_{j_1,j_2} \leq t_0 < \tilde{T}_{j_1,j_2}+\eps$. Since this holds for any $\eps > 0$, this implies $T_{j_1,j_2} \leq \tilde{T}_{j_1,j_2}$. Putting everything together, we have shown that if $A_{n+j_1,n+j_2} = 1$ and $|\hat{\mu}(X_{n+j_1}) - \hat{\mu}(X_{n+j_2})| > T_{j_1,j_2}$ hold, then $\tilde{T}_{j_1,j_2} = T_{j_1,j_2}$.

Therefore,
\begin{align*}
    &\EE{e_{j_1,j_2} \cdot \One{H_{j_1,j_2}}} = \EE{e_{j_1,j_2} \cdot A_{n+j_1,n+j_2}}\\
    &= \EE{\frac{\One{|\hat{\mu}(X_{n+j_1}) - \hat{\mu}(X_{n+j_2})| > T_{j_1,j_2}}\cdot A_{n+j_1,n+j_2}}{\sum_{i=1}^{n_1} \One{|\hat{\mu}(X_{2i-1}) - \hat{\mu}(X_{2i})| > T_{j_1,j_2}}\cdot \One{(Y_{2i-1}-Y_{2i})(\hat{\mu}(X_{2i-1}) - \hat{\mu}(X_{2i})) \le 0} + 1} \cdot (n_1+1)}\\
    &= \EE{\frac{\One{|\hat{\mu}(X_{n+j_1}) - \hat{\mu}(X_{n+j_2})| > T_{j_1,j_2}} \cdot \One{|\hat{\mu}(X_{n+j_1}) - \hat{\mu}(X_{n+j_2})| > T_{j_1,j_2}}\cdot A_{n+j_1,n+j_2}}{\sum_{i=1}^{n_1} \One{|\hat{\mu}(X_{2i-1}) - \hat{\mu}(X_{2i})| > T_{j_1,j_2}}\cdot A_{2i-1,2i} + 1} \cdot (n_1+1)}\\
    &= \EE{\frac{\One{|\hat{\mu}(X_{n+j_1}) - \hat{\mu}(X_{n+j_2})| > \tilde{T}_{j_1,j_2}} \cdot \One{|\hat{\mu}(X_{n+j_1}) - \hat{\mu}(X_{n+j_2})| > T_{j_1,j_2}}\cdot A_{n+j_1,n+j_2}}{\sum_{i=1}^{n_1} \One{|\hat{\mu}(X_{2i-1}) - \hat{\mu}(X_{2i})| > \tilde{T}_{j_1,j_2}}\cdot A_{2i-1,2i} + 1} \cdot (n_1+1)}\\
    &\leq \EE{\frac{\One{|\hat{\mu}(X_{n+j_1}) - \hat{\mu}(X_{n+j_2})| > \tilde{T}_{j_1,j_2}} \cdot A_{n+j_1,n+j_2}}{\sum_{i=1}^{n_1} \One{|\hat{\mu}(X_{2i-1}) - \hat{\mu}(X_{2i})| > \tilde{T}_{j_1,j_2}}\cdot A_{2i-1,2i} + 1} \cdot (n_1+1)}\\
    &\leq \EE{\frac{\One{|\hat{\mu}(X_{n+j_1}) - \hat{\mu}(X_{n+j_2})| > \tilde{T}_{j_1,j_2}} \cdot A_{n+j_1,n+j_2}}{\sum_{i=1}^{n_1} \One{|\hat{\mu}(X_{2i-1}) - \hat{\mu}(X_{2i})| > \tilde{T}_{j_1,j_2}}\cdot A_{2i-1,2i} + \One{|\hat{\mu}(X_{n+j_1}) - \hat{\mu}(X_{n+j_2})| > \tilde{T}_{j_1,j_2}} \cdot A_{n+j_1,n+j_2}} \cdot (n_1+1)}\\
    & \le 1,
\end{align*}
where the last inequality holds since the pairs $((X_1,Y_1),(X_2,Y_2)), ((X_3,Y_3),(X_4,Y_4)),\ldots,$ $((X_{2n_1-1},Y_{2n_1-1}),(X_{2n_1},Y_{2n_1}))$ and $((X_{n+j_1},Y_{n+j_1}),(X_{n+j_2},Y_{n+j_2}))$ are exchangeable, and $\tilde{T}_{j_1,j_2}$ is invariant under arbitrary permutations of these pairs. This completes the proof.

\subsection{Proof of Theorem~\ref{thm:fdr}}
\label{app:proof-fdr}
Let $\mathcal{R}$ denote a rejection set satisfying the 
self-consistency condition~\eqref{eq:self-consistency} at level $\alpha$.
The false discovery proportion of $\mathcal{R}$ can be expressed and bounded as
\begin{align*}
&\FDP = \frac{\sum_{1 \leq j_1 < j_2 \leq m}  \One{H_{j_1,j_2} \text{ is true}} \cdot \One{(j_1,j_2) \in \mathcal{R}}}{|\mathcal{R}| \vee 1}\\
& \le \frac{\sum_{1 \leq j_1 < j_2 \leq m}  \One{H_{j_1,j_2} \text{ is true}} \cdot \One{e_{j_1,j_2} \ge 
\frac{m(m-1)/2}{\alpha |\mathcal{R}|}}}{|\mathcal{R}| \vee 1} \leq \alpha \cdot \frac{\sum_{1 \leq j_1 < j_2 \leq m}  \One{H_{j_1,j_2} \text{ is true}} \cdot e_{j_1,j_2} }{m(m-1)/2},
\end{align*}
where the first inequality follows from the self-consistency condition~\eqref{eq:self-consistency}.
Taking expectation on both sides, we have
\begin{align*}
    \FDR = \EE{\FDP} & \leq \EE{\alpha \cdot  \frac{\sum_{1 \leq j_1 < j_2 \leq m} \One{H_{j_1,j_2} \text{ is true}} \cdot e_{j_1,j_2} }{m(m-1)/2}} \\ 
    & = \frac{\alpha}{m(m-1)/2} \cdot \sum_{1 \leq j_1 < j_2 \leq m} \EE{e_{j_1,j_2} \cdot  \One{H_{j_1,j_2} \text{ is true}}} \leq \frac{\alpha}{m(m-1)/2} \cdot \sum_{1 \leq j_1 < j_2 \leq m} 1 = \alpha,
\end{align*}
as desired, where the last inequality follows from Lemma~\ref{lem:fdr}.

\subsection{Proof of Proposition~\ref{prop:equiv}}
\label{app:proof-equiv}
We write $\mathcal{R}_{\ebh}$ as a shorthand for $\mathcal{R}_{\textnormal{e-BH}}(\{\tilde{e}_{j_1,j_2}\}_{1 \leq j_1 < j_2 \leq m})$ and let $\tilde \delta_{j_1,j_2} = \One{(j_1,j_2) \in \mathcal{R}_{\ebh}}$, for $1\le j_1 < j_2\le m$.

For any $(j_1,j_2) \in \mathcal{R}_{\ebh}$, we have $\tilde e_{j_1,j_2} \le \tilde e_{a,b}$  
for any $(a, b)$ such that $B_{j_1,j_2} \subseteq B_{a,b}$ by construction, which implies $\tilde\delta_{j_1,j_2} \le \tilde \delta_{a,b}$.
for any $(j_1', j_2')$ such that $B_{j_1,j_2} \subseteq B_{j_1',j_2'}$ by construction---which implies $\tilde\delta_{j_1,j_2} \le \tilde \delta_{j_1',j_2'}$. 
Since the e-BH procedure is self-consistent~\citep{wang2022false},  we have 
\[
\alpha e_{j_1,j_2} \sum_{1 \le a < b\le m} \tilde \delta_{a,b} \ge 
\alpha \tilde e_{j_1,j_2} \sum_{1 \le a < b\le m} \tilde \delta_{a,b} \ge \tilde \delta_{j_1,j_2}
m(m-1)/2.
\]
for any $(j_1,j_2) \in \mathcal{R}_{\ebh}$.
We have therefore verified that $\{\tilde \delta_{j_1,j_2}\}$ is a feasible solution to 
the linear program~\eqref{eq:lp}.

Assume for the purpose of contradiction that there exists a pair 
$(k_1,k_2) \in \mathcal{R}_{\ebh}$ such that $\hat \delta_{k_1,k_2} = 0$.
Define a new solution $\delta'$, where for any $1\le j_1 < j_2 \le m$, 
\begin{align*}
\delta_{j_1,j_2}' = \begin{cases}
1 & \text{ if } B_{k_1,k_2} \subseteq B_{j_1,j_2}\\ 
\hat \delta_{j_1,j_2} & \text{otherwise.}
\end{cases}
\end{align*}
We can check that $\delta_{j_1,j_2}' \le \delta_{a,b}'$, whenever $B_{j_1,j_2} \subseteq B_{a,b}$.
In addition, for $(j_1,j_2)$ such that $B_{k_1,k_2} \subseteq B_{j_1,j_2}$, 
\begin{align*}
\alpha e_{j_1,j_2}  \sum_{1 \le a < b \le m } \delta_{a,b}' \ge 
\alpha \tilde e_{k_1,k_2}  \sum_{1 \le a < b \le m } \hat \delta_{a,b}
\ge \alpha \tilde e_{k_1,k_2}  \sum_{1 \le a < b \le m } \tilde \delta_{a,b}
\ge  m(m-1)/2,
\end{align*}
where the second-to-last step is because $\sum_{1 \le a < b \le m } \hat \delta_{a,b} \ge \sum_{1 \le a < b \le m } \tilde \delta_{a,b}$, 
and the last step is because $(k_1,k_2) \in \mathcal{R}_{\ebh}$.
Also, for any $(j_1,j_2)$ such that $B_{k_1,k_2} \not\subseteq B_{j_1,j_2}$, 
\begin{align*}
\alpha e_{j_1,j_2}  \sum_{1 \le a < b \le m } \delta_{a,b}' \ge 
\alpha e_{j_1,j_2}  \sum_{1 \le a < b \le m } \hat \delta_{a,b}
\ge  \hat \delta_{j_1,j_2} m(m-1)/2,
\end{align*}
where  the last step is because $\hat \delta$ is a feasible solution. 
So far, we have verified that $\delta'$ is also a feasible solution to~\eqref{eq:lp}, 
but apparently $\sum_{j_1 < j_2} \delta'_{j_1,j_2} > \sum_{j_1 < j_2} \hat \delta_{j_1,j_2}$, 
which contradicts the optimality of $\hat \delta_{j_1,j_2}$.
As a result, we must have $\mathcal{R}_{\ebh} \subseteq \{(j_1,j_2): \hat \delta_{j_1,j_2}=1\}$.

On the other hand, consider any $(j_1,j_2)$ such that $\hat \delta_{j_1,j_2} = 1$.
For any $(j_1',j_2')$ such that $B_{j_1,j_2} \subseteq B_{j_1',j_2'}$, we have $\hat \delta_{j_1',j_2'} = 1$
by consistency and 
\begin{align*}
\alpha e_{j_1',j_2'}  \sum_{1 \le a < b \le m } \hat \delta_{a,b} \ge m(m-1)/2,
\end{align*}
which implies that $\alpha \tilde e_{j_1,j_2} \sum_{1 \le a < b \le m } \hat \delta_{a,b} \ge m(m-1)/2$.
By the form of e-BH, this implies $(j_1,j_2) \in \mathcal{R}_{\ebh}$.
Putting the two directions together, we have $\mathcal{R}_{\ebh} = \{(j_1,j_2): \hat \delta_{j_1,j_2}=1\}$.

\subsection{Proof of Theorem~\ref{thm:rank_fdr}}
\label{app:proof-rank-fdr}

Let us define $E_k$ as the set of edges in $\mathcal{R}_{\rm ord}$ with left node $k$, i.e.,
\begin{equation*}
    E_k = \{(k,l) : l > k, (k,l) \in \mathcal{R}_{\rm ord}\}.
\end{equation*}
Note that $E_1,\ldots,E_m$ form a partition of $\mathcal{R}_{\rm ord}$, i.e., the sets are disjoint and their union is $\mathcal{R}_{\rm ord}$. Let
\[
t_k = |\{(k,l) \in E_k : H_{\hat I_k\wedge \hat I_l, \hat I_k\vee \hat I_l} \textnormal{ is true}\}|,\qquad
f_k = |\{(k,l) \in E_k : H_{\hat I_k\wedge \hat I_l, \hat I_k\vee \hat I_l} \textnormal{ is false}\}|.
\]
By the definition of $\hat{u}_k = m-|E_k|$, we have $\hat{u}_k + t_k + f_k = m$ for all $k \in [m]$. Now observe that the following inequality holds for each $j \in [m]$:
\[
R_{\hat I_j}  \leq m-t_{j}
= \hat{u}_{j}+f_{j},
\]
since there are at least $t_{j}$ outcomes larger than $Y_{n+\hat I_j}$, corresponding to the correctly certified comparisons in $E_{j}$. Therefore,
\[
d_+(R_{\hat I_j},\ch_{\hat I_j}) = d_+(R_{\hat I_j},\tilde C_j) 
= (R_{\hat I_j}-\hat{u}_{j})_+
\leq f_{j}.
\]
Moreover, $m-\hat{u}_k=|E_k|$ for every $k\in[m]$. Combining these observations yields
\begin{align*}
\EE{\frac{\sum_{j=1}^m d_+(R_j,\ch_j)}
{\max\{\sum_{k=1}^m(m-\hat{u}_k),1\}}}
& = \EE{\frac{\sum_{j=1}^m d_+(R_{\hat I_j},\ch_{\hat I_j})}
{\max\{\sum_{k=1}^m |E_k|,1\}}}\\
& \leq
\EE{\frac{\sum_{j=1}^m f_{j}}
{\max\{\sum_{k=1}^m|E_k|,1\}}}
=
\EE{\frac{\sum_{k=1}^m f_k}{\max\{|\mathcal{R}|,1\}}}
=\FDR\leq\alpha.
\end{align*}

\end{document}